\documentclass[aps,superscriptaddress,nofootinbib,floatfix,mamsmath,11pt]{revtex4}
\usepackage{epsfig,epsf,graphics,psfrag}
\usepackage{times}
\usepackage{float}
\usepackage{amsfonts,amssymb,stmaryrd,latexsym,amsmath}
\usepackage{slashed}
\usepackage{bm}
\usepackage[backref=none]{hyperref} 

\newcommand{\be}{\begin{equation}}
\newcommand{\ee}{\end{equation}}
\newcommand{\ba}{\begin{eqnarray}}
\newcommand{\ea}{\end{eqnarray}}

\usepackage{url}

\begin{document}
\title{Theoretical constraints and systematic effects in the determination of the proton form factors}

\author{I.~T.~Lorenz}
\email{lorenzi@hiskp.uni-bonn.de}
\affiliation{Helmholtz-Institut f\"ur Strahlen- und
             Kernphysik and Bethe Center for Theoretical Physics, \\
             Universit\"at Bonn,  D--53115 Bonn, Germany}

\author{Ulf-G.~Mei{\ss}ner}
\email{meissner@hiskp.uni-bonn.de}
\affiliation{Helmholtz-Institut f\"ur Strahlen- und
             Kernphysik and Bethe Center for Theoretical Physics, \\
             Universit\"at Bonn,  D--53115 Bonn, Germany}
\affiliation{Institute for Advanced Simulation, Institut f\"{u}r Kernphysik
             and J\"ulich Center for Hadron Physics, \\
             Forschungszentrum J\"{u}lich, D--52425 J\"{u}lich, Germany}

\author{H.-W.~Hammer}
\affiliation{Institut f\"ur Kernphysik, Technische Universit\"at Darmstadt, 64289 Darmstadt, Germany}
\affiliation{ExtreMe Matter Institute EMMI, GSI Helmholtzzentrum f\"ur
Schwerionenforschung GmbH, 64291 Darmstadt, Germany}

\author{Y.-B.~Dong}
\affiliation{Institute of High Energy Physics, Beijing 100049, People's Republic of China}
\affiliation{Theoretical Physics Center for Science Facilities (TPCSF), CAS, Beijing 100049, People's Republic of China}

\begin{abstract}
\noindent We calculate the two-photon exchange corrections to electron-proton scattering with nucleon and $\Delta$ intermediate states. The results show a dependence on the elastic nucleon and nucleon-$\Delta$-transition form factors used as input which leads to significant changes compared to previous calculations. We discuss the relevance of these corrections and apply them to the most recent and precise data set and world data from electron-proton scattering. Using this, we show how the form factor extraction from these data is influenced by the subsequent inclusion of physical constraints. The determination of the proton charge radius from scattering data is shown to be dominated by the enforcement of a realistic spectral function. Additionally, the third Zemach moment from the resulting form factors is calculated. The obtained radius and Zemach moment are shown to be consistent with Lamb shift measurements in muonic hydrogen.
\end{abstract}


\maketitle

\section{Introduction}
\begin{figure}[t]
\centering
\includegraphics[width=0.25\textwidth]{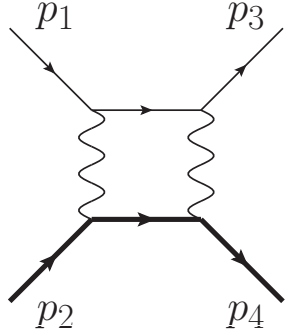}
\caption{Box graph, calculated here with different form factor parametrizations, crossed box implied.\label{fig:box}}
\end{figure}

\noindent 
The basic building block of visible matter, the nucleon, and its electromagnetic interaction lie at the heart of several precise QED calculations. One of those refers to the Lamb shift, the splitting between the $j=1/2$ 2$S$ and 2$P$ levels of hydrogen. The measurement of this energy splitting initially triggered the development of quantum field theory since this shift goes beyond the prediction of the Dirac equation. Nowadays, the Lamb shift can be measured not only in regular electronic hydrogen but also in muonic hydrogen. However, the results are inconsistent when using the same proton structure information for both \cite{Pohl:2010zza}. The required proton structure information is partly encoded in the electromagnetic nucleon form factors (NFFs), most relevant are the gradients at the origin, defining the radius. The Lamb shift inconsistency can be expressed via the electric proton radius $r_E^p$. Furthermore, recent measurements of the NFFs from electron-proton scattering have been used to determine $r_E^p$ with a claimed accuracy that excludes the value from the muonic Lamb shift \cite{Bernauer10, Bernauer:2013tpr}. However, previous physically motivated fits to the same data found perfect agreement with the latter value \cite{Lorenz:2012tm} but were criticized due to small systematic deviations from the data. Besides possibly neglected experimental systematics, radiative corrections could also explain such deviations in principle. Earlier physically constrained fits to the world NFF data found similar values for $r_E^p$ \cite{Belushkin:2006qa} in agreement with the result from muonic hydrogen.\newline
Particularly interesting in this regard are corrections with intermediate resonant states that have not been included in the standard corrections to elastic electron-proton (e-p) scattering. The largest contribution of this kind is expected from the graph with two photons (see Fig.~\ref{fig:box}) and the first excited state of the nucleon, the $\Delta$ resonance, since it has the lowest mass and the strongest nucleon coupling of such resonances. Cosmologically, it is largely responsible for the Greisen-Zatsepin-Kuzmin (GZK) cutoff which limits the energy of cosmic ray protons via their interaction with photons from the microwave background \cite{Greisen:1966jv}. Phenomenologically, information on this resonance comes mainly from photo- and electroproduction processes, but it can also be seen in neutrino reactions. Such information including the momentum dependence of the vertices \cite{Pascalutsa:2006up, Tiator:2003uu} is used in this work to calculate the given correction. The calculation requires a treatment in the Rarita-Schwinger formalism for spin-$3/2$ \cite{Rarita:1941mf}.\newline
The other main interest of this paper lies in the theoretical constraints that one can impose on suitable NFF parametrizations. An example of an analytically reasonable parametrization is based on a conformal mapping of the domain of analyticity onto the unit circle and a subsequent expansion, often denoted by $z$. This is widely used for transition form factors in heavy meson decays, see e.g. \cite{Okubo:1971jf, Bourrely:1980gp, Boyd:1994tt, Abbas:2009dz, Ananthanarayan:2011uc}. In such works, unitarity constraints from the physical pair-production region are also often introduced, based on an operator product expansion. For the NFFs, a continuum contribution due to $2\pi$ exchange occurs far below the pair-production threshold, but close to the physical scattering region. The spectral function corresponding to this continuum and higher pole contributions can be determined from related processes. Loose constraints from the spectral function on coefficients in a $z$ expansion have been introduced in a recent analysis \cite{Hill:2010yb}. The complete known information on this spectral function can be included via dispersion relations. We examine in this paper whether loose constraints are sufficient to obey the tight constraint on the nucleon spectral functions set by unitarity.\newline 
We perform a complete analysis of the current world cross sections on elastic electron-proton cross sections in order to extract the NFFs. Here, we examine the interplay of statistical and systematic effects with the inclusion of the theoretical constraints from analyticity and unitarity. 
The inclusion of the complete data range also allows us to extract the ``third Zemach moment,'' which is relevant in the calculation of the Lamb shift, especially in muonic hydrogen \cite{Friar:2005jz}. 
The particular pursuit of accuracy in this work is needed for subsequent work on the NFFs for timelike momentum transfer. These can be related to the spacelike form factors via the inclusion of weight functions with as particular analytic behavior in dispersion relations \cite{Geshkenbein:1974gm}.\newline
The paper is structured as follows. In Sec. II of this work, we explicitly calculate the TPE corrections including nucleon and $\Delta$ intermediate states and realistic vertices. In Sec. III, we discuss in detail the theoretical constraints that can be imposed on the NFFs and perform fits to TPE-corrected cross sections. The extracted NFFs are used in Sec. IV to determine the third Zemach moment. We conclude with a discussion in Sec. V. Further details on fits and statistics are given in the Appendices.

\section{Form factor extraction and corrections}
\subsection{Definitions of form factors and helicity amplitudes}
\noindent First, we consider the hadronic matrix element of the electromagnetic current for the nucleon ground state. The helicity can either be conserved or flipped, which is parametrized in the common separation into the Dirac FF, $F_1$, and the Pauli FF, $F_2$,
\begin{align}\label{nvertex}
 \langle N(p')|J^{\mu}_{em}|N(p)\rangle = ie\bar{u}(p')\left(\gamma^{\mu}F_1(t)+i\frac{\sigma^{\mu\nu}q_{\nu}}{2m_N}F_2(t)\right)u(p) = ie\bar{u}(p')\Gamma^{\mu}(t)u(p),
\end{align}
where $t = q^2 = (p'-p)^2 = -Q^2$ is the invariant momentum transfer squared, and $m_N$ is the nucleon mass. For electron-nucleon scattering, we have $Q^2\geq 0$. $F_1^{p/n}(0)$ and $F_2^{p/n}(0)$ are given in terms of the proton/neutron electric charge and anomalous magnetic moment, respectively. The separation of these form factors in their isoscalar and the isovector parts is given by $ F_i^{s} = (F_i^p+F_i^n)/2$ and $F_i^{v} = (F_i^p-F_i^n)/2$ for $i =1,2$, correspondingly. In order to avoid interference terms, the cross section is often considered in a different FF basis, the electric and magnetic Sachs form factors $G_{E,M}^{p,n}(t)$
\begin{align}
 G_E(t) &= F_1(t)-\tau F_2(t),\notag\\
 G_M(t) &= F_1(t)+F_2(t),
\end{align}
with $\tau = -t/4m_N^2$. At first order in the fine-structure constant $\alpha$, the Born-approximation, the differential cross section can be expressed through these Sachs FFs as
\begin{equation}\label{eq:xs_ros}
\frac{d\sigma}{d\Omega} = \left( \frac{d\sigma}
{d\Omega}\right)_{\rm Mott} \frac{\tau}{\epsilon (1+\tau)}
\left[G_{M}^{2}(Q^{2}) + \frac{\epsilon}{\tau} G_{E}^{2}(Q^{2})\right]\, ,
\end{equation}
where $\epsilon = [1+2(1+\tau)\tan^{2} (\theta/2)]^{-1}$ is the virtual photon polarization, $\theta$ is the electron scattering angle in the
laboratory frame, and $({d\sigma}/{d\Omega})_{\rm Mott}$ is the Mott cross section, which corresponds to scattering off a pointlike particle. Two quantities out of energies, momenta and angles suffice to determine this cross section and are related for such an elastic process. Specifically, in the laboratory frame with the initial nucleon at rest and neglecting the electron mass, we can write
\begin{align}
 Q^2 \approx 4E_1E_3\sin^2\left(\frac{\theta}{2}\right),
\end{align}
where $E_1(E_3)$ are the energies of the incoming (outgoing) electron.\newline 
Moreover, we are interested here in the transition from the nucleon to the $\Delta$ resonance with the corresponding matrix element,
\begin{align}
 \langle \Delta(p')|J^{\nu}_{em}|N(p) \rangle = \Psi_{\mu}(p')\Gamma^{\mu\nu}_{\gamma N\rightarrow\Delta}(p',q)u(p) 
\end{align}
where in addition to the usual Dirac spinor $u(p)$ we introduce the Rarita-Schwinger spinor field $\Psi_{\mu}^{(a)}(p)$ \cite{Rarita:1941mf}. Due to the spin-$3/2$ nature of the $\Delta$, this has two indices, the Lorentz index $\mu$ and the spinor index $(a)$, which is neglected in the formulas for clarity. As for the NFFs, the first suggestions for N$\Delta$-vertex decompositions were constructed in a way that diagonalizes the cross section \cite{Jones:1972ky}. However, for a comparison to previous similar calculations in the literature, we consider here a parametrization given in Ref.~\cite{Kondratyuk:2001qu}
\begin{align}\label{dvertex}
 \Gamma_{\gamma N\rightarrow\Delta}^{\mu\nu}(p',q) = \frac{ieF_{dip}(q^2)}{2m_{\Delta}^2}\{g_1[g^{\nu\mu}\slashed{p}'\slashed{q} - p'^{\nu}\gamma^{\mu}\slashed{q} - \gamma^{\nu}\gamma^{\mu}p'\cdot q + \gamma^{\nu}\slashed{p}'q^{\mu}] + g_2[p'^{\nu}q^{\mu} - g^{\nu\mu}p'\cdot q]\notag\\
 + g_3/m_{\Delta}[q^2(p'^{\nu}\gamma^{\mu} - g^{\nu\mu}\slashed{p}') + q^{\nu}(q^{\mu}\slashed{p}' - \gamma^{\mu}p'\cdot q)]\}\gamma_5T_3.
\end{align}
where $T_3$ is the isospin transition factor, $p'$ the four-momentum of the outgoing $\Delta$ and $q$ of the incoming photon. Linear combinations of $g_1, g_2$ and $g_3$ describe the magnetic, electric and Coulomb parts $g_M$, $g_E$ and $g_C$ of the transition, respectively, that can be obtained from experiment for vanishing $q^2$. Here, a dipole-behavior $F_{dip} = \Lambda^4/(\Lambda^2-q^2)^2$ of the $q^2$ dependence is assumed, with $\Lambda = 0.84$~GeV. The inclusion of a more realistic $q^2$ dependence is possible with a reformulation via helicity amplitudes, since these can be given from resonance-electroproduction data analyses. For the introduction of these, we stay close to Ref.~\cite{Piranishvili:2008zz, Lalakulich:2006sw}. In order to relate their set of form factors $C_3(q^2), C_4(q^2)$ and $C_5(q^2)$ to the helicity amplitudes, we rewrite the vertex as 
\begin{align}
 \langle \Delta(p')|J^{em}\cdot\epsilon|N(p) \rangle &= \Psi_{\mu}(p')\Gamma_{\nu}^C F^{\mu\nu}u(p)\notag\\
 &= \Psi_{\mu}(p')\left(\frac{C_3(q^2)}{m_N}\gamma_{\nu}+\frac{C_4(q^2)}{m_N^2}p'_{\nu}+\frac{C_5(q^2)}{m_N^2}p_{\nu}\right)\gamma_5(q^{\mu}\epsilon^{\nu}-q^{\nu}\epsilon^{\mu})u(p).
\end{align}
Here, the polarization vector of the (virtual) photon $\epsilon^{\mu}$ can correspond to a right-/left-handed transverse or a longitudinal polarization. Using again the reference frame with the initial nucleon at rest and the photon in $z$-direction, the polarization vectors can be written as
\begin{align}
 \epsilon^{\mu(R/L)} = \mp\frac{1}{\sqrt{2}}(0,1,\pm i,0), \hspace{8pt} \epsilon^{\mu(S)} = \frac{1}{\sqrt{Q^2}}(q_3,0,0,q_0).
\end{align}
These polarizations induce the possible transitions of the helicity $\lambda$ in the hadron (R) states $|R, \lambda\rangle$: 
\begin{align}
 A_{1/2}(Q^2) = N \langle \Delta, +\frac{1}{2}|J^{em}_{\mu}\cdot \epsilon^{\mu(R)}|N,-\frac{1}{2}\rangle\xi,\\
 A_{3/2}(Q^2) = N \langle \Delta, +\frac{3}{2}|J^{em}_{\mu}\cdot \epsilon^{\mu(R)}|N,+\frac{1}{2}\rangle\xi,\\
 S_{1/2}(Q^2) = N \frac{q_3}{Q^2} \langle \Delta, +\frac{1}{2}|J^{em}_{\mu}\cdot \epsilon^{\mu(S)}|N,+\frac{1}{2}\rangle\xi.\label{helamp}
\end{align}
where $N = \pm\sqrt{\pi\alpha/m_N(W^2-m_N^2)}$ and the phase $\xi$ is determined empirically here. The transverse~(T) and longitudinal~(L) parts of the cross section of the electroproduction of the $\Delta$ at its mass can be given in terms of these helicity amplitudes as
\begin{align}
 \sigma_T(W = m_{\Delta}) &= \frac{2m_N}{m_{\Delta}\Gamma_{\Delta}}(A^2_{1/2}(Q^2) + A^2_{3/2}(Q^2)),\notag\\ 
 \sigma_L(W = m_{\Delta}) &= \frac{2m_N}{m_{\Delta}\Gamma_{\Delta}}\frac{Q^2}{q_3^2}S_{1/2}^2(Q^2).
\end{align}
The approximations inherent in the Breit-Wigner-definitions of mass $m_{\Delta}$ and width $\Gamma_{\Delta}$ are expected to be small compared to the remaining uncertainties. For the $\Delta$ resonance this approximation is more reasonable than for higher partial waves. For higher resonances, masses and widths should always be taken from the pole position obtained in a dynamical coupled-channel approach, see for example Ref.~\cite{Ronchen:2012eg}. 

\subsection{Two-photon exchange corrections}
\noindent The corrections to the electron-proton cross sections at order $\alpha^3$ are given by the interference of the one-photon-exchange amplitude $\mathcal{M}_{1\gamma}$ and the amplitudes from vacuum polarization, vertex corrections, self-energy corrections and the two-photon-exchange amplitude $\mathcal{M}_{2\gamma}$ and additionally the contribution from Bremsstrahlung. The main data set that we will consider in this work already contains a set of calculations of such corrections by Maximon and Tjon \cite{Maximon:2000hm}. This calculation contains improvements towards earlier works by Mo and Tsai \cite{MT69} but still uses a soft-photon approximation, particularly relevant for the two-photon-exchange (TPE) contribution. This contribution to the corrected cross section can be expressed through a factor of $(1 + \delta_{2\gamma})$ as
\begin{align}
 \frac{d\sigma_{corr.}}{d\Omega} &= (\mathcal{M}^\dagger_{1\gamma} + \mathcal{M}^\dagger_{2\gamma} + \ldots) (\mathcal{M}_{1\gamma} + \mathcal{M}_{2\gamma} + \ldots) = \frac{d\sigma_{1\gamma}}{d\Omega}(1 + \delta_{2\gamma} + \ldots),\notag\\
 &\Rightarrow \delta_{2\gamma} \underbrace{\approx}_\text{$\mathcal{O}(\alpha)$} \frac{2\text{Re}(\mathcal{M}^{\dagger}_{1\gamma} \mathcal{M}_{2\gamma})}{|\mathcal{M}_{1\gamma}|^2}.\label{tpe}
\end{align}
We briefly discuss the soft-photon approximation by Maximon and Tjon since only the difference between a new evaluation and this approximation is required for the data. They separate the IR-divergent part of the TPE amplitude by considering the poles in the photon propagators, i.e. one vanishing photon momentum. The resulting factor is 
\begin{align}
 \delta_{2\gamma, \text{IR}}^{\text{Max.-Tjon}} = -\frac{2\alpha}{\pi}\ln\frac{E_1}{E_3}\ln\frac{Q^2}{\lambda^2}
\end{align}
where $\lambda$ is an infinitesimal photon mass and $E_1~(E_3)$ again the incoming~(outgoing) electron energy. The logarithmic infrared singularity in $\lambda$ is canceled by a term in the Bremsstrahlung correction, so that the complete cross section is $\lambda$-independent. The same cancellation takes place, if both $\delta_{2\gamma, \text{IR}}$ and the Bremsstrahlung correction are calculated in the older approximation scheme by Mo and Tsai.
The hard two-photon corrections without excitations of the intermediate states have been calculated by Blunden $\it{et~al.}$ \cite{BMT05}. However, these are not included in the supplied e-p scattering cross sections. Thus these calculations are carried out here for the kinematics required by the data. Since some of the data contain the TPE calculation by Maximon and Tjon and some that by Mo and Tsai, we calculate the difference to these approximations. This calculation serves as a cross-check and starting point for the $\Delta$ TPE. Also for an excited intermediate $\Delta$ resonance there exist approximate calculations \cite{Kondratyuk:2005kk, Graczyk:2013pca, Zhou:2014xka}, but without realistic information on all vertices, including the $Q^2$ dependence. Thus we improve upon these calculations and apply them to the required cross-section kinematics. The general structures that we consider in the following are as in Ref.~\cite{Kondratyuk:2005kk} the interference between the $1\gamma$ amplitude
\begin{align}
 \mathcal{M}_{1\gamma} = -\frac{e^2}{q^2}\bar{u}_e(p_3)\gamma_{\mu}u_e(p_1)\bar{u}_N(p_4)\Gamma^{\nu}u_N(p_2)
\end{align}
and the $2\gamma$ amplitude
\begin{align*}
 \mathcal{M}_{2\gamma}^{\rm box} = -ie^4 \int \frac{d^4k}{(2\pi)^4}L_{\mu\nu}^{\rm box}H_{N/\Delta}^{\mu\nu}D(k)D(q-k).
\end{align*}
In this notation, the metric tensor from the photon propagator has already been contracted and the leptonic tensor for the box and crossed box graph, respectively, is given by
\begin{align*}
 L_{\mu\nu}^{\rm box} = \bar{u}_e(p_3)\gamma_{\mu}S_F(p_1-k,m_e)\gamma_{\nu}u_e(p_1);\hspace{5pt} L_{\mu\nu}^{\rm xbox} = \bar{u}_e(p_3)\gamma_{\nu}S_F(p_3+k,m_e)\gamma_{\mu}u_e(p_1),
\end{align*}
whereas the hadronic tensor for nucleon or $\Delta$ intermediate states are
\begin{align}
 H_N^{\mu\nu} &= \bar{u}_N(p_4)\Gamma^{\mu}(q-k)S_F(p_2+k,m_N)\Gamma^{\nu}(k)u_N(p_2)~~~~~{\rm and}\notag\\
 H_{\Delta}^{\mu\nu} &= \bar{u}_N(p_4)\Gamma_{\gamma\Delta\rightarrow N}^{\mu\alpha}(p_2+k,q-k)S_{\alpha\beta}(p_2+k)\Gamma_{\gamma N\rightarrow\Delta}^{\beta\nu}(p_2+k,k)u_N(p_2),
\end{align}
respectively. $\Gamma^{\mu}(q)$ is the elastic nucleon vertex from Eq.~\eqref{nvertex} and $\Gamma_{\gamma\Delta\rightarrow N}^{\mu\alpha}(p,k)$ the transition vertex from Eq.~\eqref{dvertex}. One can write $\Gamma_{\gamma\Delta\rightarrow N}^{\mu\alpha}(p,k) = \gamma_0[\Gamma_{\gamma N\rightarrow\Delta}^{\alpha\mu}(p,k)]^{\dagger}\gamma_0$ if one considers the momenta of $\Delta$ and photon in the conjugated vertex reversed to the original one, as in Ref.~\cite{Arrington:2011dn}, for a discussion see also Ref.~\cite{Zhou:2009nf}. In the denominator of the photon propagator for the pure nucleon graph, we include an infinitesimal photon mass $\lambda$
\begin{align}
 D(k) = \frac{1}{k^2-\lambda^2+i\epsilon},
\end{align}
to regulate the infrared divergences. The loop containing the $\Delta$ is not IR divergent because of the mass of the $\Delta$.\newline
The propagators of the nucleon and the electron have the usual form
\begin{align}
 S_F(k,m) = \frac{\slashed{k}+m}{k^2-m^2+i\epsilon}.
\end{align}
For the case of the $\Delta$, the propagator has the structure $-S_F(p_{\Delta},m_{\Delta})\mathcal{P}_{\alpha\beta}(p_{\Delta})$ with the projector for the spin-$3/2$ components
\begin{align}
 \mathcal{P}_{\alpha\beta}(p_{\Delta}) = g_{\alpha\beta}-\frac{1}{3}\gamma_{\alpha}\gamma_{\beta}-\frac{1}{3p_{\Delta}^2}(\slashed{p}_{\Delta}\gamma_{\alpha}p_{\Delta\beta} + \gamma_{\beta}\slashed{p}_{\Delta}p_{\Delta\alpha}).
\end{align}
In both the nucleon- and $\Delta$-TPE graphs, ultraviolet divergences are suppressed by the momentum dependence of the form factors.\newline
Numerically, the form factors appearing at the photon-baryon vertices 
are handled analogously to the denominators of the propagators. Therefore, the 
integrals take the structure of 4-point functions through replacements 
of the general form
\begin{align}
 \frac{1}{(\Lambda^2-k^2)k^2}=\frac{1}{\Lambda^2}\left(\frac{1}{k^2}
+\frac{1}{\Lambda^2 - k^2}\right).
\end{align}

\subsection{Results on two-photon-exchange calculations}
\noindent We use two independent symbolic manipulation programs, FORM \cite{Vermaseren:2000nd} and FeynCalc \cite{Mertig:1990an} for the analytic replacements and the trace algebra. We further reduce the integral expressions to scalar Passarino-Veltman integrals \cite{Passarino:1978jh} which are well known and can subsequently be evaluated via the program LoopTools \cite{Hahn:1998yk}.\newline
\subsubsection{Two-photon exchange: Nucleon intermediate states}
\begin{figure}[t]
\centering
\includegraphics[width=0.35\textwidth, angle=270]{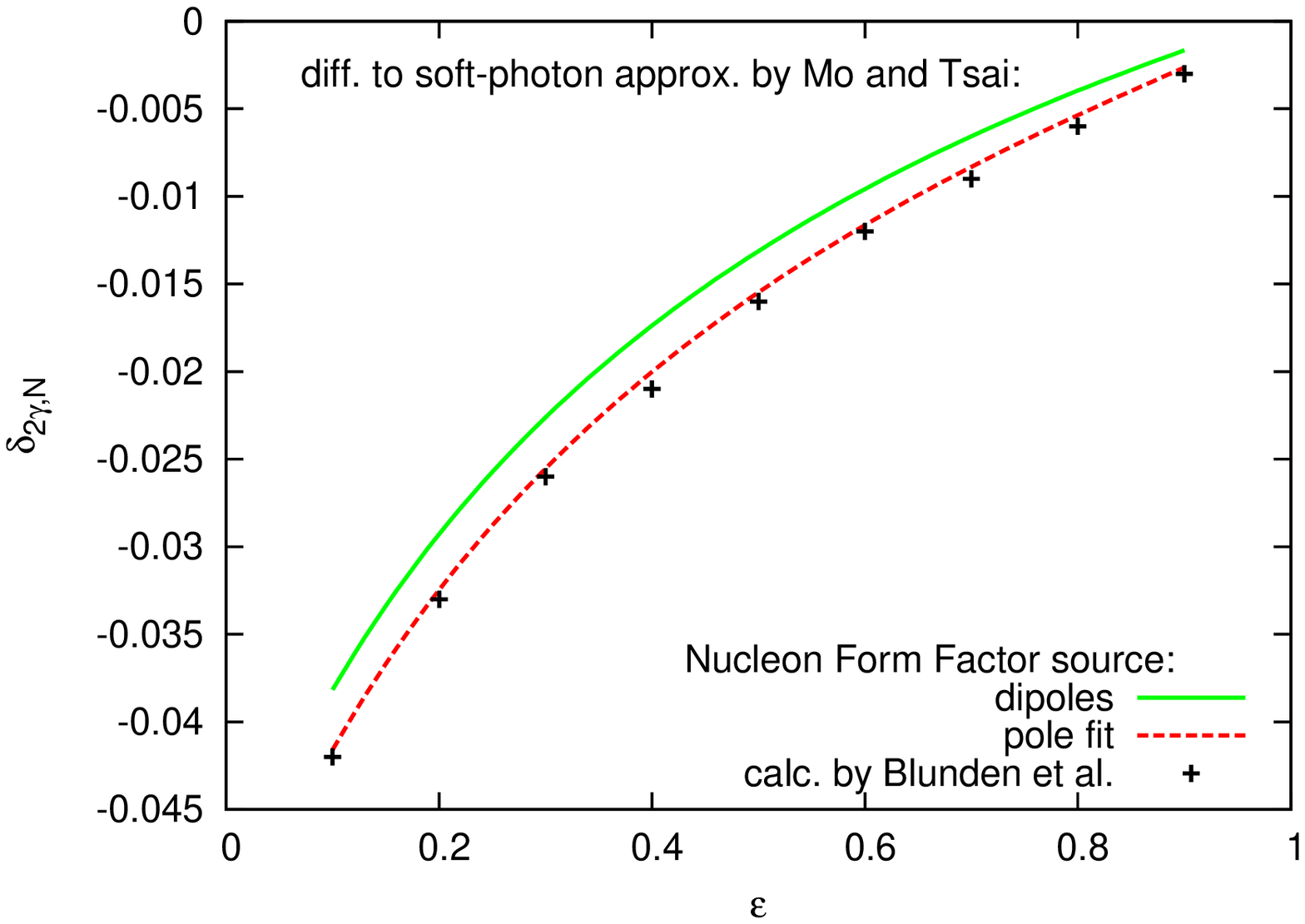}\hglue1mm
\includegraphics[width=0.35\textwidth, angle=270]{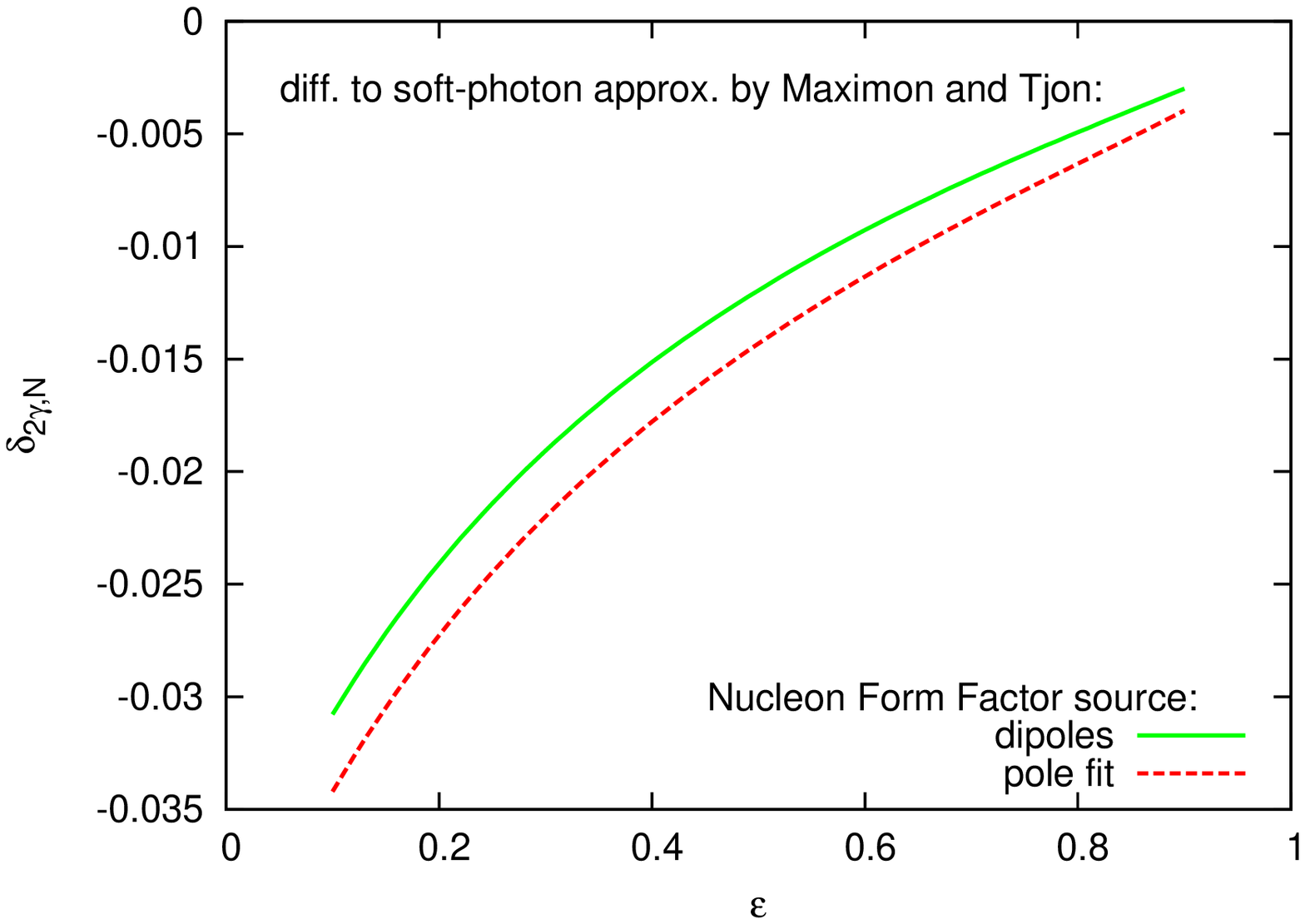}
\caption{Dependence of the TPE with nucleon intermediate state on the nucleon form factors at $Q^2$ = 3~GeV$^2$. The correction factor $\delta_{2\gamma,N}$ is calculated once with dipole Sachs FFs and once with the simplified pole fit from Ref.~\cite{BMT05}. Left panel: Difference of our calculation to the soft-photon approximation by Mo and Tsai \cite{MT69}. Right panel: Difference of our calculation to the soft-photon approximation by Maximon and Tjon \cite{Maximon:2000hm}.\label{fig:nuc}}
\end{figure}

\noindent We calculate the differences to the two soft-photon approximations contained in the different cross sections. In Fig.~\ref{fig:nuc}, we show the $\epsilon$ dependence at $Q^2$ = 3\,GeV$^2$ which allows us to compare to previous calculations by Blunden $\it{et~al.}$ In this case, we can confirm that the dependence on the NFF parametrization largely cancels out. The use of the pole fit parametrization from Ref.~\cite{BMT05} here indeed reproduces their result. Lowering the $Q^2$ value in the calculation decreases the nucleon-TPE correction. 

\subsubsection{Two-photon exchange: $\Delta$ intermediate states}
\begin{figure}[t]
\centering
\includegraphics[width=0.35\textwidth, angle=270]{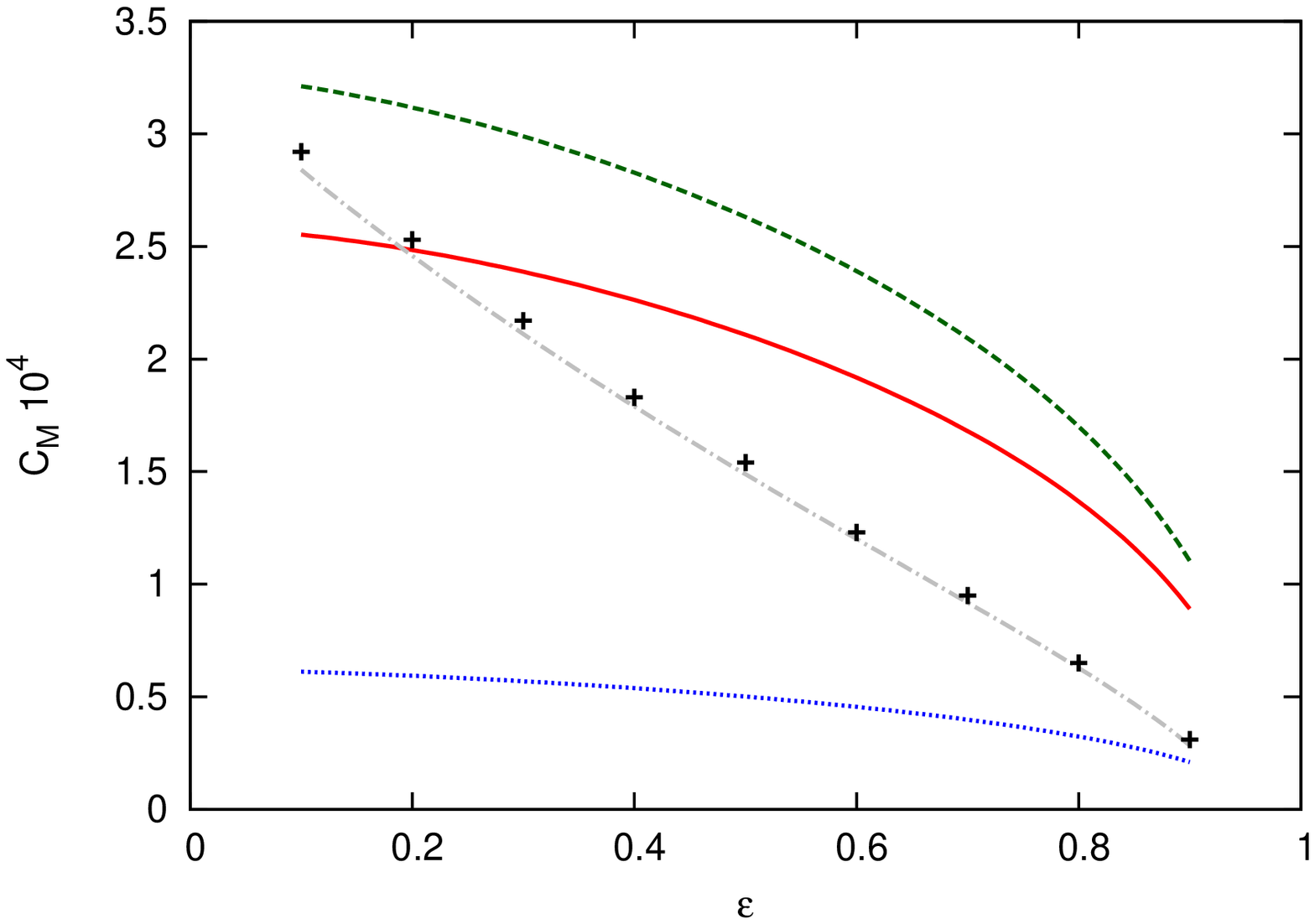}\hglue1mm
\includegraphics[width=0.35\textwidth, angle=270]{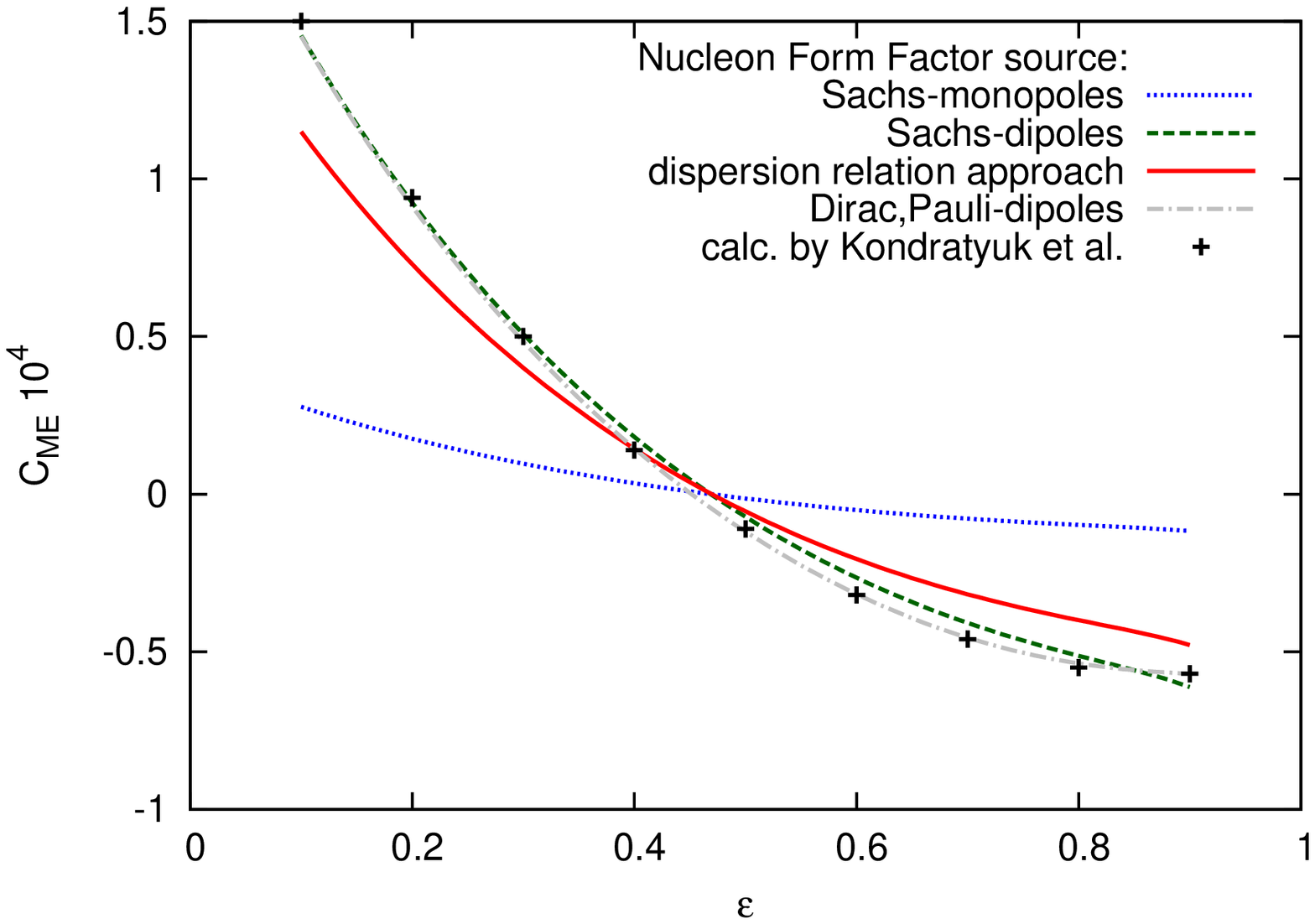}\vglue1mm
\includegraphics[width=0.35\textwidth, angle=270]{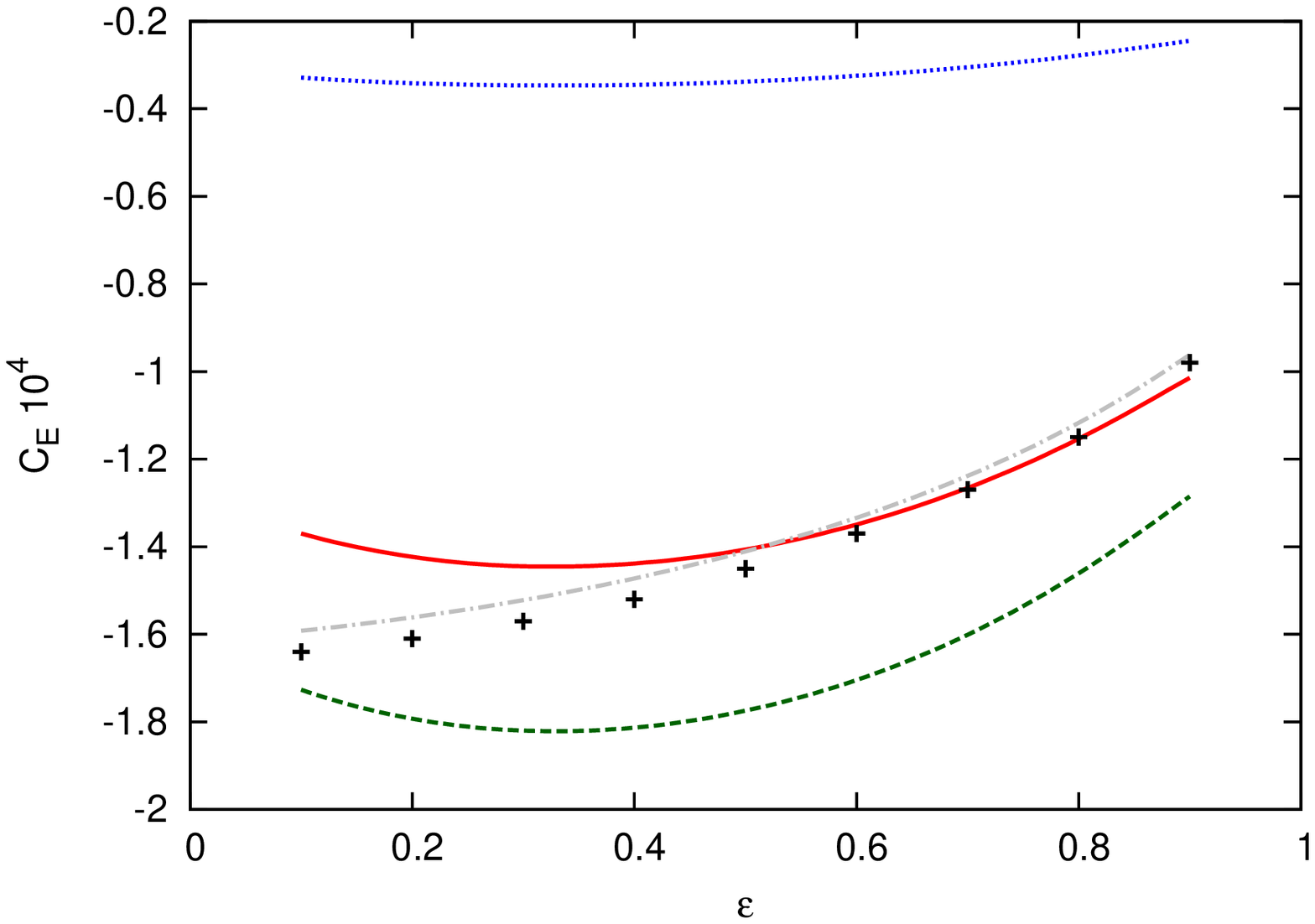}\hglue1mm
\includegraphics[width=0.35\textwidth, angle=270]{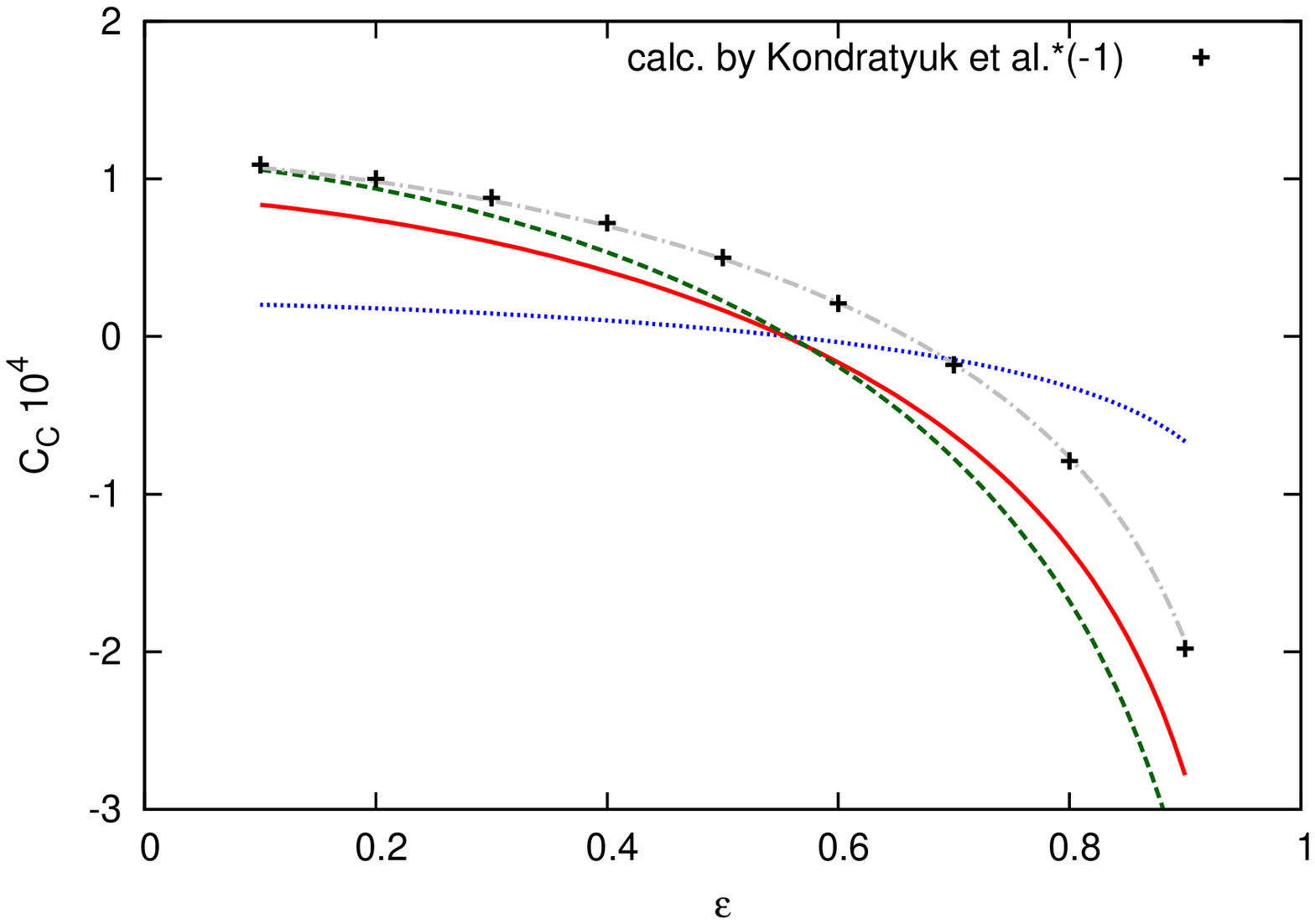}
\caption{Individual contributions to TPE with $\Delta$ intermediate state at $Q^2$ = 3~GeV$^2$.\label{fig:indiv}}
\end{figure}

\begin{figure}[t]
\centering
\includegraphics[width=0.35\textwidth, angle=270]{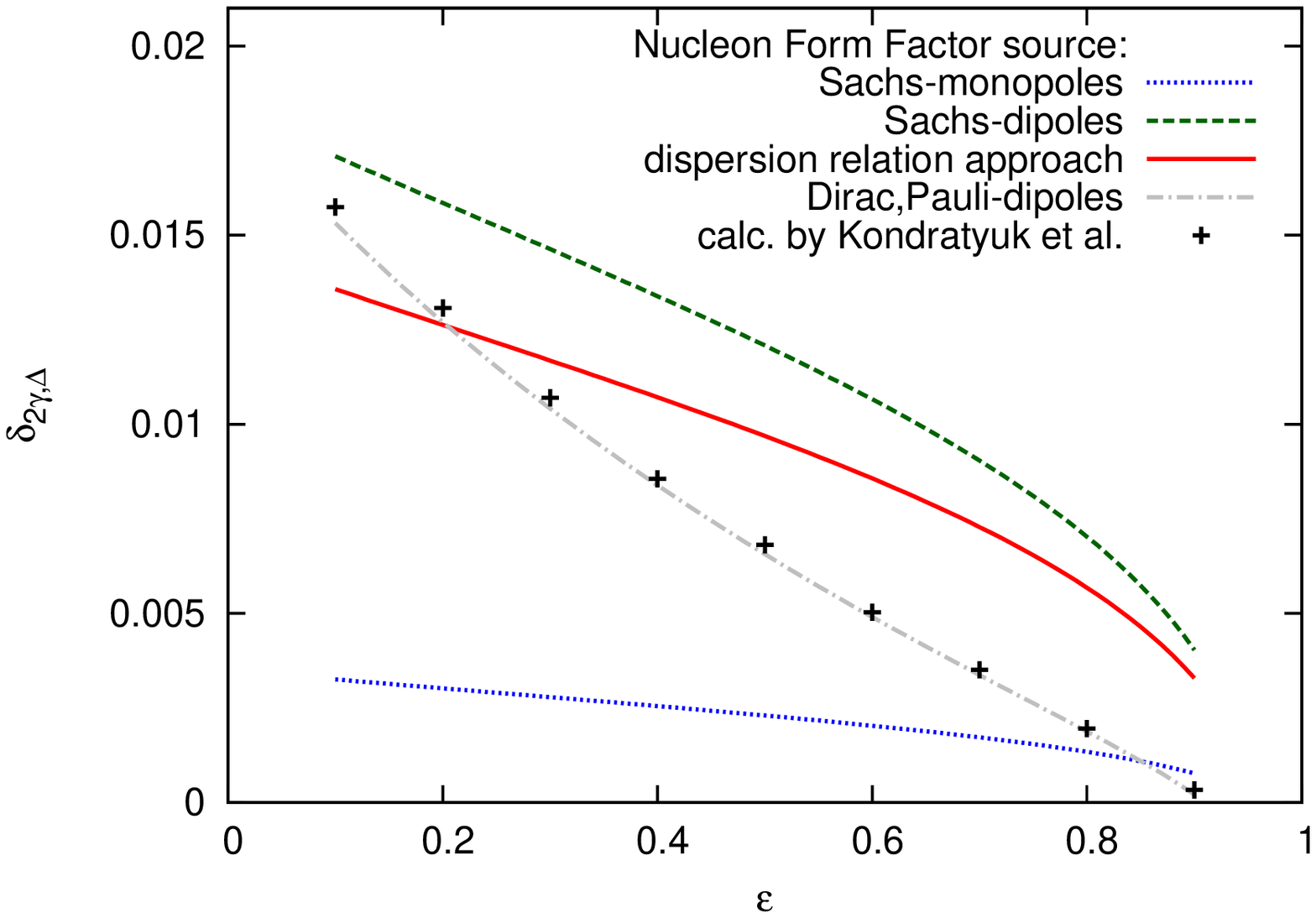}\hglue1mm
\includegraphics[width=0.35\textwidth, angle=270]{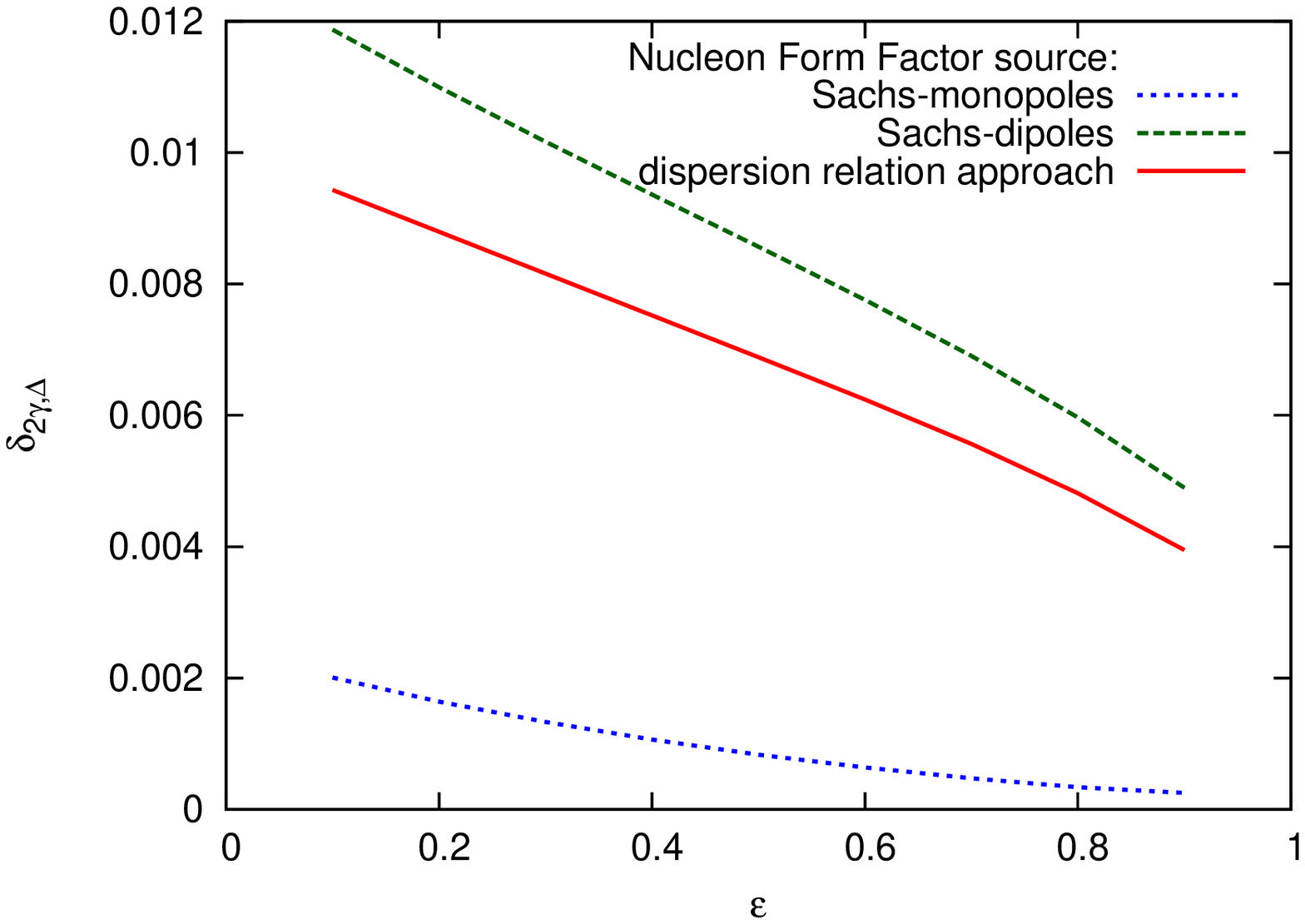}
\caption{Dependence of the TPE with $\Delta$ intermediate state on the nucleon form factors at $Q^2$ = 3~GeV$^2$. Left panel: $N\Delta\gamma$ vertex as given by Kondratyuk. Right panel: $N\Delta\gamma$ vertex directly matched to helicity amplitudes from electroproduction of nucleon resonances.\label{fig:del}}
\end{figure}

\noindent For the $\Delta$ intermediate state, the situation is different. In order to identify the main sources of uncertainty for the $\Delta$ TPE, we perform two different procedures for the treatment of the nucleon-$\Delta$ transition and vary the NFF input for both of them. First, we consider the $\gamma N\Delta$ vertex from Eq.~\eqref{dvertex} and calculate the individual contributions to the correction factor Eq.~\eqref{tpe} of the form
\begin{align}
 \delta_{2\gamma,\Delta} = C_M^{}g_M^2 + C_E^{}g_E^2 + C_C^{}g_C^2 
+ C_{ME}^{}g_E^{}g_M^{} + C_{CM}^{}g_C^{}g_M^{} + C_{CE}^{}g_C^{}g_E^{}~,\label{indiv}
\end{align}
with $g_M=g_1$, $g_E=g_2-g_1$ and $g_C=g_3$. In this case, the $Q^2$ dependence of the $\gamma N\Delta$ vertex is assumed to follow a dipole behavior and only the photocouplings $g_M, g_E, g_C$ at $Q^2=0$ are fixed to the experimental values. This approach is again directly comparable to an older calculation, by Kondratyuk $\it{et~al.}$, if we use the same values for $g_1, g_2, g_3$ as 7,9 and 0, respectively. Recent values of 6.59, 9.08 and 7.12 as used in \cite{Zhou:2014xka} taken from \cite{Pascalutsa:2006up}, change the corrections less than the NFF variations shown in Fig.~\ref{fig:indiv}. This plot shows the contributions to $\delta_{2\gamma,\Delta}$ as defined in Eq.~\eqref{indiv} for different parametrizations of the NFFs used in the $1\gamma$ amplitude of the interference term. First, we tried also here to reproduce an older calculation by Kondratyuk $\it{et~al.}$ by using the FFs quoted in their paper \cite{Kondratyuk:2005kk}. This attempt failed, and we tested other NFFs as input, e.g. the dipole parametrization of the Sachs FFs as a reasonable first approximation. Looking for possible reasons for the deviation, we also considered the NFFs used in a later paper by Kondratyuk $\it{et~al.}$ \cite{Kondratyuk:2006ig} on $\Delta$ production. Instead of the Sachs form factors, in that work, the Dirac and Pauli FFs are parametrized as dipole FFs, which contradicts empirical as well as theoretical information. However, the TPE calculation based on this agrees with the Kondratyuk calculation, as shown in Fig.~\ref{fig:indiv} (grey, dash-dotted). The small deviations here are of the order of changes due to the numerical precision of constants required in the TPE integrals. We also show the large range that is covered by using a dipole (green, long-dashed) or monopole (blue, short-dashed) for the Sachs FFs. The red (solid) curve shows the TPE contribution calculated with the most realistic FFs, from our dispersion relation fit, albeit one to not fully corrected data \cite{Lorenz:2012tm}. The numerical values corresponding to the DR-FFs in Fig.~\ref{fig:indiv} are tabulated in Tab.~\ref{table:corr}. The dominating magnetic contribution shows the largest deviation from Kondratyuk $\it{et~al.}$, if we neglect the sign change in the Coulomb-contribution. As discussed in Ref.~\cite{Zhou:2014xka}, a sign change here can be related to the correct inclusion of the photon momentum direction. For the comparison to the Kondratyuk calculation, this sign change has no impact on the whole correction $\delta_{2\gamma,\Delta}$ due to their approximation $g_C=0$. This complete correction is shown in the left panel of Fig.~\ref{fig:del}. The mixing terms of the Coulomb- with other contributions are $\leq10^{-10}$ for our DR-FF input.\newline
\begin{table}[ht!]
 \centering
\begin{tabular}{|l|c|c|c|c|}
\hline
$\epsilon$ & $C_M\times10^4$ & $C_{ME}\times10^4$ & $C_E\times10^4$ & $C_C\times10^4$\\
\hline
$0.1$ & $2.55$ & $1.15$ & $-1.37$ & $0.84$ \\
$0.2$ & $2.48$ & $0.73$ & $-1.42$ & $0.74$ \\
$0.3$ & $2.39$ & $0.40$ & $-1.44$ & $0.60$ \\
$0.4$ & $2.26$ & $0.14$ & $-1.44$ & $0.41$ \\
$0.5$ & $2.11$ & $-0.05$ & $-1.41$ & $0.17$ \\
$0.6$ & $1.92$ & $-0.21$ & $-1.35$ & $-0.16$ \\
$0.7$ & $1.68$ & $-0.32$ & $-1.27$ & $-0.63$ \\
$0.8$ & $1.37$ & $-0.40$ & $-1.15$ & $-1.35$ \\
$0.9$ & $0.89$ & $-0.48$ & $-1.01$ & $-2.78$ \\
\hline
\end{tabular}
\caption{The $\epsilon$-dependence of the coefficients $C_M,C_{ME},C_E,C_C$ for $Q^2$ = 3~GeV$^2$ and the NFF parametrization from \cite{Lorenz:2012tm}.}
\label{table:corr}
\end{table}
\noindent The right panel of Fig.~\ref{fig:del} shows the results of our second calculation that employs different information on the $\gamma N\Delta$-vertices to examine the uncertainties. We consider here the helicity amplitudes Eq.~\eqref{helamp} obtained from data on electroproduction of nucleon resonances \cite{Tiator:2003uu}. These can be parametrized conveniently by a set of FFs, determined in Ref.~\cite{Lalakulich:2006sw} and used in Ref.~\cite{Graczyk:2013pca} for a similar calculation albeit without realistic NFFs. This form of the $\gamma N\Delta$ vertex does not deviate significantly from recent data and is numerically well treatable. Even though the curvature in the $\epsilon$ dependence of the correction changes slightly when the helicity amplitudes are used, the relative change in magnitude at different kinematics is smaller than the NFF dependence. For most of the given kinematics, we can see the calculation in the left panel as an upper limit.\newline 
The comparison with Ref.~\cite{Zhou:2014xka} is reassuring: they employ different transition form factors in the main part, however, with a consistent treatment of the four-momentum signs in a Kondratyuk-like calculation, they obtain the same sign change for the Coulomb contribution as we do. Moreover, their shown result from such a calculation with $g_C = 0$ also deviates from the original paper \cite{Kondratyuk:2005kk} but the NFF dependence is not further studied or discussed at all. Further calculations for different kinematics and assumptions are not directly comparable \cite{Tomalak:2014sva, Gorchtein:2014hla, Borisyuk:2013hja, Afanasev:2005mp}.\newline

\subsubsection{Application to cross sections}
\begin{figure}[t]
\centering
\includegraphics[width=0.6\textwidth]{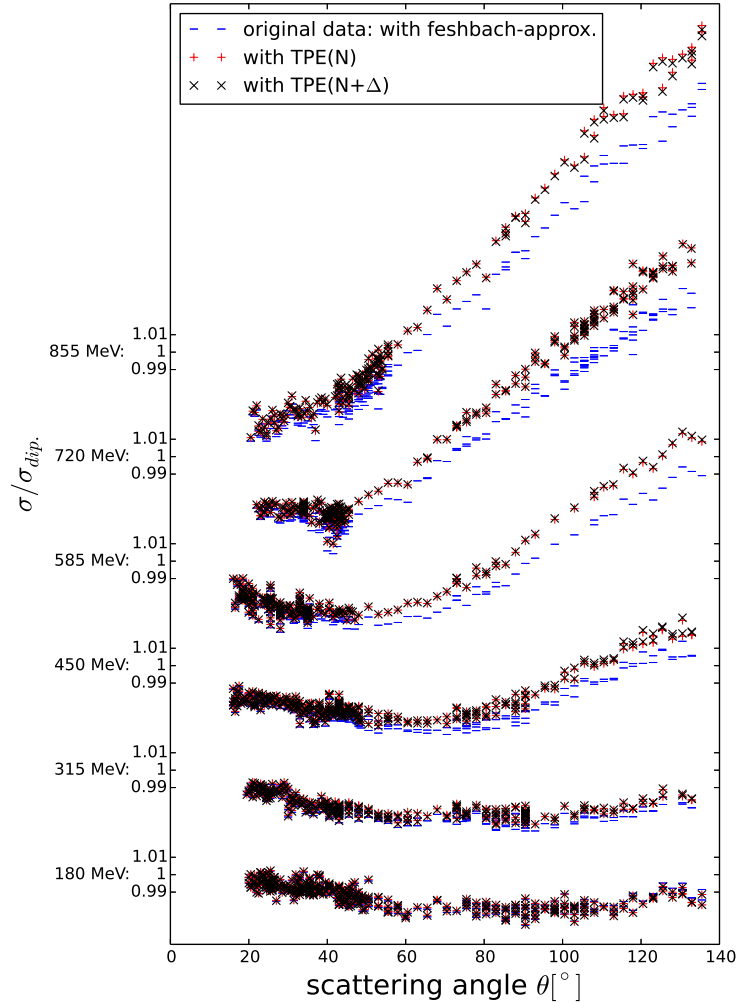}
\caption{The impact of TPE corrections on the electron-proton scattering cross sections with the highest quoted precision \cite{Bernauer10}. From the original data, we subtract the Feshbach approximation and add our calculations. We display the cross section divided by that one calculated by dipole Sachs FFs to make the deviations clearer.\label{fig:app}}
\end{figure}

\noindent We apply our calculated corrections to the electron-proton scattering data with the highest quoted precision at low $Q^2$. The corresponding measurements have been carried out at the Mainz Microtron (MAMI) for six different energies of the incoming electron beam with three spectrometers by the A1 collaboration \cite{Bernauer10, Bernauer:2013tpr}. We display the cross sections with an offset for these six energy settings in Fig.~\ref{fig:app} depending on the scattering angle $\theta$. The original data contains an approximation of the two-photon correction that is only valid in the limit $Q^2\rightarrow 0$, which even has the wrong sign for some kinematical regions, as shown by Arrington \cite{Arrington:2011kv}. This approximation is given in the simple form
\begin{equation}
 \delta_F = Z\alpha\pi\frac{\sin\frac{\theta}{2}-\sin^2\frac{\theta}{2}}{\cos^2\frac{\theta}{2}} 
\end{equation}
by Feshbach and Kinley \cite{McKinley:1948zz}. We subtract this and replace it by our calculations. For the nucleon intermediate state, we have seen that the dependence on the nucleon form factors is small at low $Q^2$ and thus we use a simple pole fit for the nucleon form factors in these calculations. For the correction from the $\Delta$ intermediate state we employ here the $\gamma N\Delta$ vertex from Eq.~\eqref{dvertex} with recent values on the photocouplings $g_1=6.59, g_2=9.08, g_3=7.12$. This serves here as an upper limit for the correction compared to the calculation based on the helicity amplitudes. Since in this case the dependence on the NFFs in the $1\gamma$ amplitude is also significant, see Fig.~\ref{fig:del}, we use those from a previous dispersion relation fit here. Besides the original MAMI cross sections we show in Fig.~\ref{fig:app} the same data corrected by our nucleon-TPE calculation (red, +) and the nucleon+$\Delta$-TPE calculation (black, x). Here, we omit the error bars to show the corrections more clearly. $Q^2$ remains below 1~GeV$^2$ for the shown MAMI data.\newline
Besides the last MAMI data set with the highest precision, we partly include in the following analysis former world data on electron-proton scattering. First, this serves as a consistency check and second, for an evaluation of the proton structure dependence of the third Zemach moment (see below) a larger data range is needed. Care has been taken about the treatment of the IR divergences. The MAMI data set contains the IR-approximation by Maximon and Tjon, the world data compilation by I. Sick \cite{Sick14} contains the one by Mo and Tsai. 

\section{Theoretically constrained fit functions}\label{theo}
\begin{figure}[t]
\centering
\includegraphics[width=0.68\textwidth]{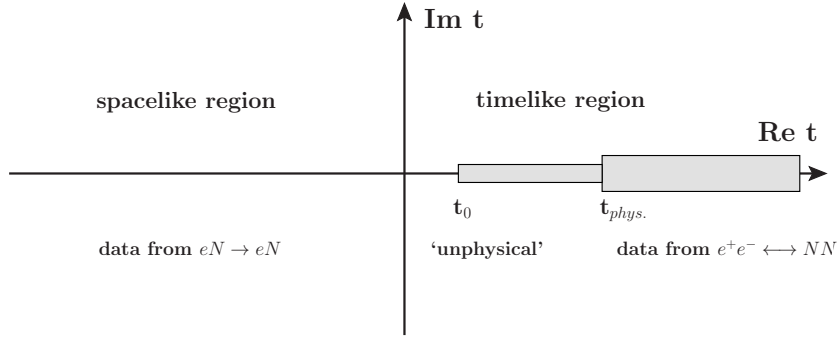}
\caption{The analytic structure of the FFs: shown is the continuation of $t = q^2$ into the complex plane, where the FFs are analytic functions except for the cut on the real axis $t>4M_{\pi}^2$. The physical FFs from scattering are defined on the negative real axis, those from creation/annihilation ($t>t_{phys.}=4m_N^2$) just above the real axis as $F(t'=t+i\epsilon)$ for infinitesimal $\epsilon$. This $\epsilon$-prescription also holds in the region $4M_{\pi}^2<t<4m_N^2$. \label{fig:spacetime}}
\end{figure}

\noindent In this section, we introduce the relevant analytic structure of the nucleon form factors and the known information on the spectral function. We point out two distinct procedures based on analyticity and unitarity to constrain the FFs via the physical and unphysical region of timelike momentum transfer, see Fig.~\ref{fig:spacetime}. We show which input has the largest impact on the FFs in the spacelike region. Based on this reasoning, we provide the FF parametrizations used in this work. 
\subsection{Analytic structure and spectral decomposition of the form factors}
\noindent For timelike momentum transfer, the NFFs are defined via the matrix element
\begin{align}
 I_{\mu} = \langle N(p)\bar{N}(\bar{p})|j_{\mu}^{em}(0)|0\rangle = \bar{u}(p)\left[\gamma_{\mu}F_1(t)+i\frac{\sigma_{\mu\nu}q^{\nu}}{2m_N}F_2(t)\right]v(\bar{p}).\label{20}
\end{align}
The insertion of a complete set of intermediate states $|\lambda\rangle$ yields the imaginary part of the FFs, the spectral function: 
\begin{align}
 \text{Im}~I_{\mu} \propto \sum_{\lambda}\langle N(p)|\bar{J}_N(0)|\lambda\rangle\langle\lambda|j_{\mu}^{em}(0)|0\rangle v(\bar{p})\delta^4(p+\bar{p}-p_{\lambda}),\label{21}
\end{align}
for more details, see Ref.~\cite{Belushkin:2006qa}. By using an unsubtracted dispersion relation (DR) for the FFs
\begin{equation}
 F_i(t) = \frac{1}{\pi}\int_{t_0}^{\infty}\frac{\text{Im}F_i(t')dt'}{t'-t}~,\hspace{1cm}i = 1,2;
\label{disprel}
\end{equation}
with $t_0 = 4M_\pi^2 \, (9M_\pi^2)$ the isovector (isoscalar) threshold, we relate the spectral function in the timelike region to the FFs in the spacelike region. Eq.~\eqref{disprel} shows directly that the lowest mass states are the most relevant.\newline
The lowest-lying state $|\lambda\rangle = |2\pi\rangle$ leads to the continuum in the isovector $(v)$ channel 
\begin{align}
 \text{Im} G_{E,v}^{(2\pi)} = \frac{q_t^3}{m_N\sqrt{t}}F_{\pi}(t)^*f^1_+(t),\label{2pi}\\
 \text{Im} G_{M,v}^{(2\pi)} = \frac{q_t^3}{\sqrt{2t}}F_{\pi}(t)^*f^1_-(t).\notag
\end{align} 
where $F_{\pi}$ is the pion vector form factor, $f^1_{\pm}$ are the analytically continued pion-nucleon p-wave helicity scattering amplitudes and $q_t$ is the pion momentum. Up to $40M_{\pi}^2$, these expressions are well known and include the $\rho$ meson as intermediate state. Information on $F_{\pi}$ can be taken from Refs.~\cite{kloe11, babar09}, on $f^1_{\pm}$ from \cite{Ditsche:2012fv}. In the isoscalar channel, the lightest vector mesons are $\omega$ and $\phi$. Their widths are negligible compared to the isovector continuum. Therefore, the corresponding spectral function can be well approximated by $\delta$ distributions. A summary of the masses and widths of the lightest vector mesons is given in Tab.~\ref{table:mesons}.\newline
Specifically, we consider parametrizations for the $2\pi$, $K\bar{K}$, and $\rho\pi$ continuum, as obtained or updated in Refs.~\cite{Lorenz:2012tm, Hammer:1999uf, Meissner:1997qt}. The $\delta$ distributions from the narrow vector mesons yield pole terms. We obtain for the complete isoscalar and -vector parts of the Dirac and Pauli form factors, respectively, 
\begin{align}
 F_i^s (t) &= \sum_{V=K\bar{K},\rho\pi,s_1,s_2,..}
\frac{a_i^V}{m_V^2-t}~,\nonumber\\
 F_i^v (t) &= \sum_{V=v_1,v_2,..}\frac{a_i^V}{m_V^2-t} +
 \frac{a_i+b_i(1-t/c_i)^{-2}}{2(1-t/d_i)}~,\label{VMDspec}
\end{align}
with $i = 1,2$. The last term in the isovector form factor corresponds to the parametrization of the two-pion continuum with values that we updated in Ref.~\cite{Lorenz:2012tm}. For the light isoscalar vector mesons, the residua in the pole terms can be related to their couplings. Only rough estimates exist for these: $0.5\,\mbox{GeV}^2<|a_{1}^\omega| < 1\,\mbox{GeV}^2$, $|a_{2}^\omega| < 0.5\,\mbox{GeV}^2$~\cite{Grein:1977mn} and $|a_{1}^\phi| < 2\,\mbox{GeV}^2$, $|a_{2}^\phi| < 1\,\mbox{GeV}^2$~\cite{Meissner:1997qt}.\newline
We want to emphasize here that the spectral function in both isospin channels is well known up to at least $40M_{\pi}^2$. Additional continua due to higher numbers of pions are strongly suppressed, as has been calculated in Chiral Perturbation Theory \cite{Bernard:1996cc}.
\begin{table}[ht!]
 \centering
\begin{tabular}{|l|c|c||l|c|c|}
\hline
isoscalar ($I^G = 0^-$)	& mass & width & isovector ($I^G = 1^+$) & mass & width\\
\hline
$\omega(782)$	&  782.65	&	0.00849	& $\rho(770)$& 775.26	& see $2\pi$-cont.\\
$\phi(1020)$	& 1019.461	& 	0.00427	& $\rho(1450)$& 1465	& 0.4	\\
$\omega(1420)$	& 1400-1450	& 	0.215	& $\rho(1700)$& 1720	& 0.25	\\
$\omega(1650)$	& 1670 		& 	0.315	&		&	&	\\
$\phi(1680)$	& 1680 		&	0.150	&		&	&	\\
$\phi(2170)$	& 2175 		&	0.061 	&		&	&	\\
\hline
\end{tabular}
\caption{Overview: Masses and widths of vector mesons with $J^{PC}=1^{--}$ in both isospin channels, in MeV ~\cite{Agashe:2014kda}.}
\label{table:mesons}
\end{table}

\subsection{Conformal mappings and related constraints}
\noindent A procedure to deal with constraints from analyticity and unitarity via the physical region has been proposed by Okubo in 1971 for the example of Kaon decays \cite{Okubo:1971jf, Okubo:1971my}. This is based on a conformal mapping which has also been considered explicitly for NFFs in order to facilitate numerical procedures at that time \cite{Hohler:1976ax}. Following Ref.~\cite{violini}, we write a function that maps the cut in the $t$-plane onto the unit circle in a new variable $z$:
\begin{equation}
 z(t,t_{\rm cut})=\frac{\sqrt{t_{\rm cut}-t}-\sqrt{t_{\rm cut}}}{\sqrt{t_{\rm
       cut}-t}+\sqrt{t_{\rm cut}}}~,\label{simple}
\end{equation}
where $t_{\rm cut} = 4M_{\pi}^2$ is the lowest singularity of the form factors with $M_{\pi}$ the charged pion mass. This allows us to expand for example the Sachs form factors in the new variable $z$:
\begin{equation}
 G_{E/M}(z(t))=\sum_{k=0}^{k_{\rm max}}e_kz(t)^k~.\label{expa}
\end{equation}
The first coefficients are determined by the form factor normalizations to the charge and anomalous magnetic moment of the proton, respectively. For the remaining coefficients, one can motivate bounds, as suggested in Refs.~\cite{Hill:2010yb, Epstein:2014zua}. They parameterize the unit circle by $z(t)=e^{i\theta(t)}$ and integrate over it. For Eq.~\eqref{simple}, this leads to the following expression for the coefficients:
\begin{align*}
 e_{k\geq1}=\frac{2}{\pi}\int_{t_{cut}}^{\infty}\frac{dt}{t} \sqrt{\frac{t_{cut}}{t-t_{cut}}} \text{Im}G(t)\sin[k\theta(t)]
\end{align*}
Inserting for the spectral function the $\delta$-distribution terms given in the last section and partly also the two-pion continuum, one finds the specific bounds on the absolute values of the coefficients $|e_k|<10$. Equivalently, such bounds can be given for the individual isospin channels.\newline
For completeness, we comment here on the procedure to include constraints from analyticity and unitarity in Kaon decays, based on Okubo's ideas \cite{Okubo:1971jf, Okubo:1971my}. These ideas were used and improved in several analyses also concerning heavy meson decays, see for example \cite{Bourrely:1980gp, Boyd:1994tt, Boyd:1997qw, Becher:2005bg}. The main ingredient in this method is the two-point correlation function of the respective current
\begin{align}
 \Pi_{\mu\nu}(q^2) = i\int d^4xe^{iq\cdot x}\langle 0|T{j_{\mu}(x)j_{\nu}(0)}|0\rangle = \frac{1}{q^2}(q_{\mu}q_{\nu}-q^2g_{\mu\nu})\Pi^T(q^2) + \frac{q_{\mu}q_{\nu}}{q^2}\Pi^L(q^2)
\end{align}
that defines the transverse and longitudinal polarization functions $\Pi^{T/L}$, respectively. On the one hand, these can be approximated for large spacelike momentum transfer by an operator product expansion (OPE). Moreover, since they are analytic below the respective threshold, they satisfy a dispersion relation
\begin{align}
 \frac{1}{n!}\frac{d^n\Pi^{L/T}(q^2)}{dq^{2n}}\bigg|_{q^2=0} = \frac{1}{\pi}\int_0^{\infty}dt \frac{\text{Im}~\Pi^{L/T}(t)}{(t-q^2)^{n+1}}\bigg|_{q^2=0},
\end{align}
where a sufficiently high number of subtractions $n$ is required for a finite dispersion relation.\newline
On the other hand, by unitarity, we can express the imaginary part of the polarization functions by inserting the sum over all allowed intermediate states $Y$
\begin{align}
 \text{Im}~\Pi^{L/T}(q^2) = \frac{1}{2}\sum_{Y}\int d\rho_Y(2\pi)^4\delta^4(q-p_Y)P_{L/T}^{\mu\nu}\langle0|j_{\mu}|Y\rangle \langle Y|j_{\nu}^{\dagger}|0\rangle \geq \text{Im}~\Pi^{L/T}_Y(q^2),
\end{align}
where $d\rho_Y$ is the phase space weighting and $P_{L/T}^{\mu\nu}$ the longitudinal/transverse helicity projector. Defining $\text{Im}~\Pi^{L/T}_Y$ as the part due to only specific intermediate states, this clearly never exceeds the complete expression. For the case of electromagnetic NFFs, it has been shown \cite{Hill:2010yb} that to first order in the OPE the contribution from the physical region is small compared to that of the unphysical region confirming the results from earlier dispersion analyses, for example Ref.~\cite{Lorenz:2012tm}.\newline
For the discussion of the convergence of a $z$-expansion approach e.g. for the pion FF in the timelike region, see Ref.~\cite{Buck:1998kp}. 

\subsection{Fit results}
\begin{figure}[t]
\centering
\includegraphics[width=0.6\textwidth]{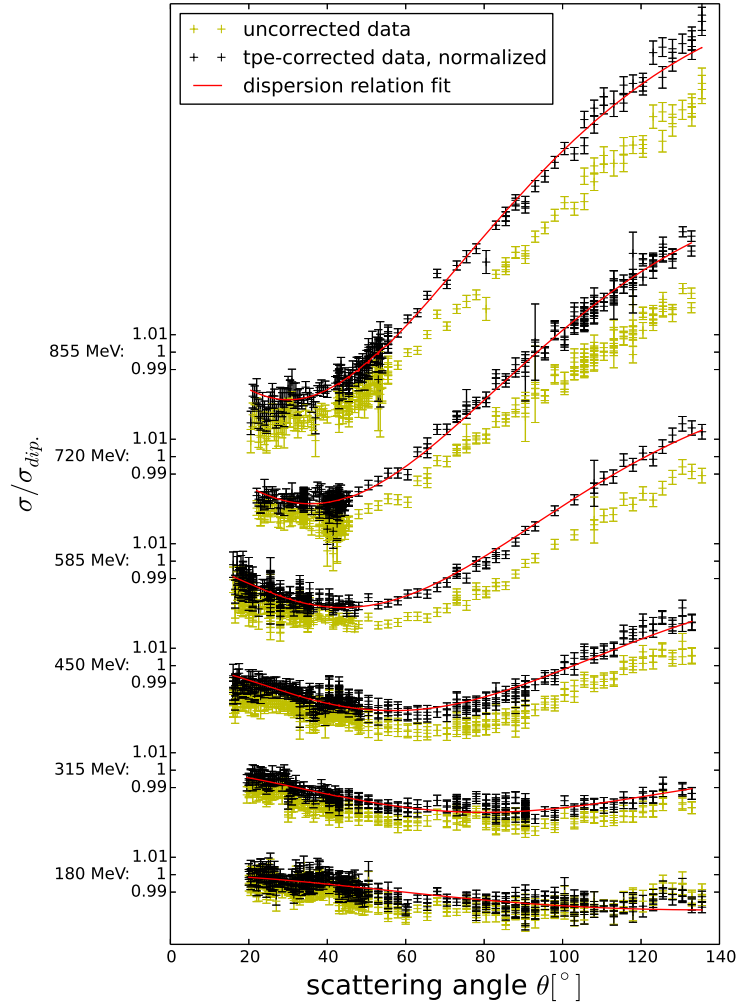}
\caption{The DR approach as given in Tab.~\ref{table:drcvalues}.\label{fig:drc}}
\end{figure}

\begin{figure}[t]
\centering
\includegraphics[width=0.5\textwidth]{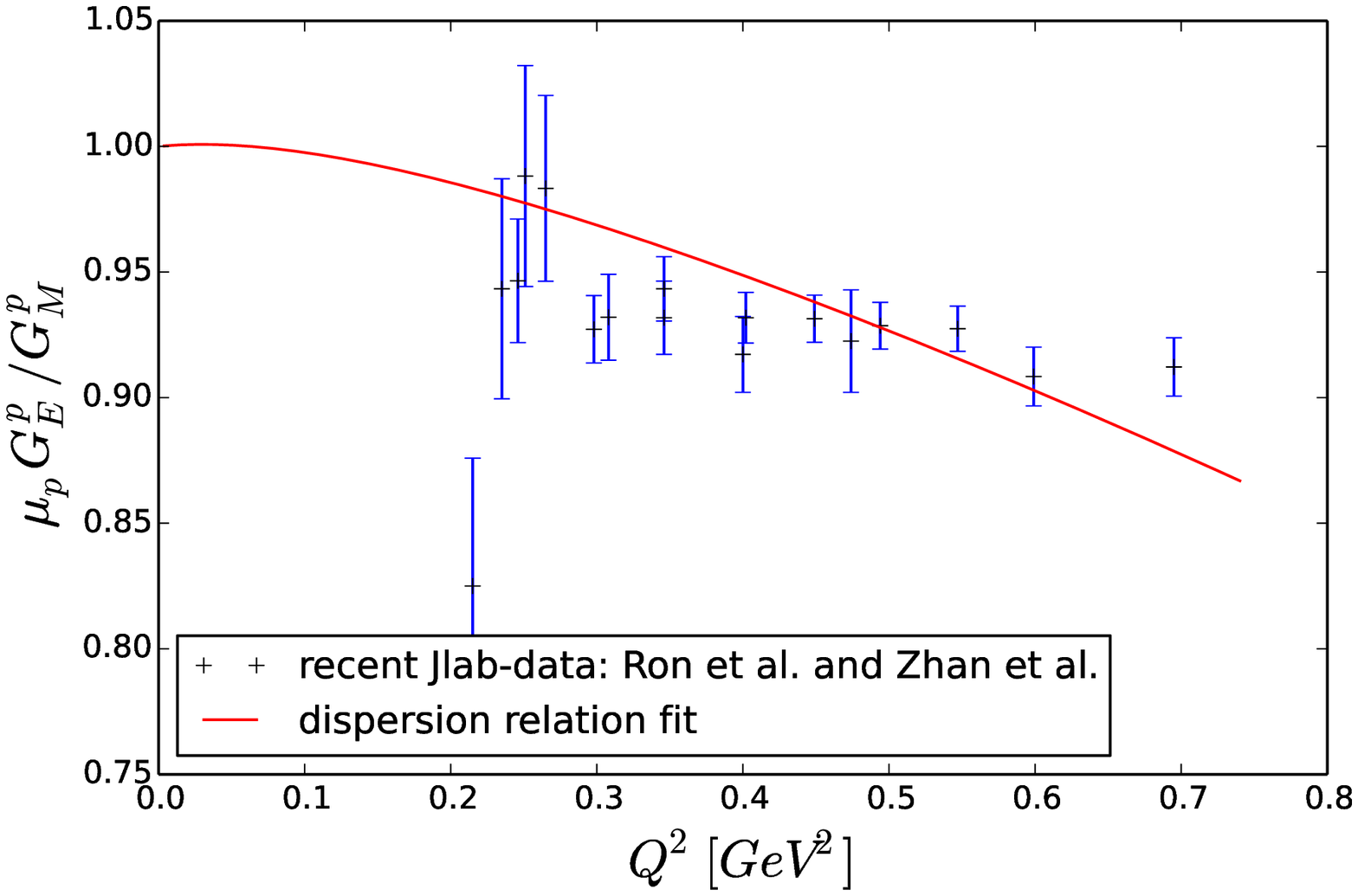}
\caption{Result from the DR approach as given in Tab.~\ref{table:drcvalues}: The prediction for the form factor ratio, compared to polarization measurements \cite{Ron:2011rd, Zhan:2011ji} that are not included in the fit here.
\label{fig:droff}}
\end{figure}

\noindent In this section, we show the results of fits to electron-proton scattering cross sections. We employ different form factor parametrizations, partly with high flexibility and partly including theoretical constraints, and fit them to different data sets. Here, we discuss the treatment of statistical and systematical uncertainties, as well as the impact of different data and corrections. The variations in the FF parametrizations allow us to analyse the influence of theoretical constraints and their explicit manifestation in the spectral function. We give the detailed results of the best physically-motivated fit in terms of the cross section, the form factor ratio, parameters and an error analysis for the radii.\newline
\begin{table}[ht!]
 \centering
\begin{tabular}{|l|c|c|}
\hline
parametrization & MAMI~(1422 data points) &  world data incl. MAMI~(1922 data points) \\
\hline
unconstrained $z$ expansion & $r_E=0.64, r_M=1.97, (\chi^2_r=1.12)$ & $r_E=0.85, r_M=0.98, (\chi^2_r=1.17)$ \\
$z$ expansion, $|e_k|<10$ & $r_E=0.91, r_M=0.79, (\chi^2_r=1.17)$ & $r_E=0.89, r_M=0.77, (\chi^2_r=1.23)$ \\
DR approach & $r_E=0.84, r_M=0.85, (\chi^2_r=1.41)$ & $r_E=0.84, r_M=0.85, (\chi^2_r=1.32)$ \\
combination of the above & $r_E=0.84, r_M=0.85, (\chi^2_r=1.38)$ & $r_E=0.84, r_M=0.85, (\chi^2_r=1.30)$ \\
\hline
\end{tabular}
\caption{Radius values $r_E^p$ and $r_M^p$ in fm from fits with different parametrizations and data sets, with the corresponding $\chi^2$/ndf. The world data contains the basis from Ref.~\cite{Sick14} and the MAMI data \cite{Bernauer10}.}
\label{table:radii}
\end{table}
\begin{figure}[t]
\centering
\includegraphics[width=0.5\textwidth]{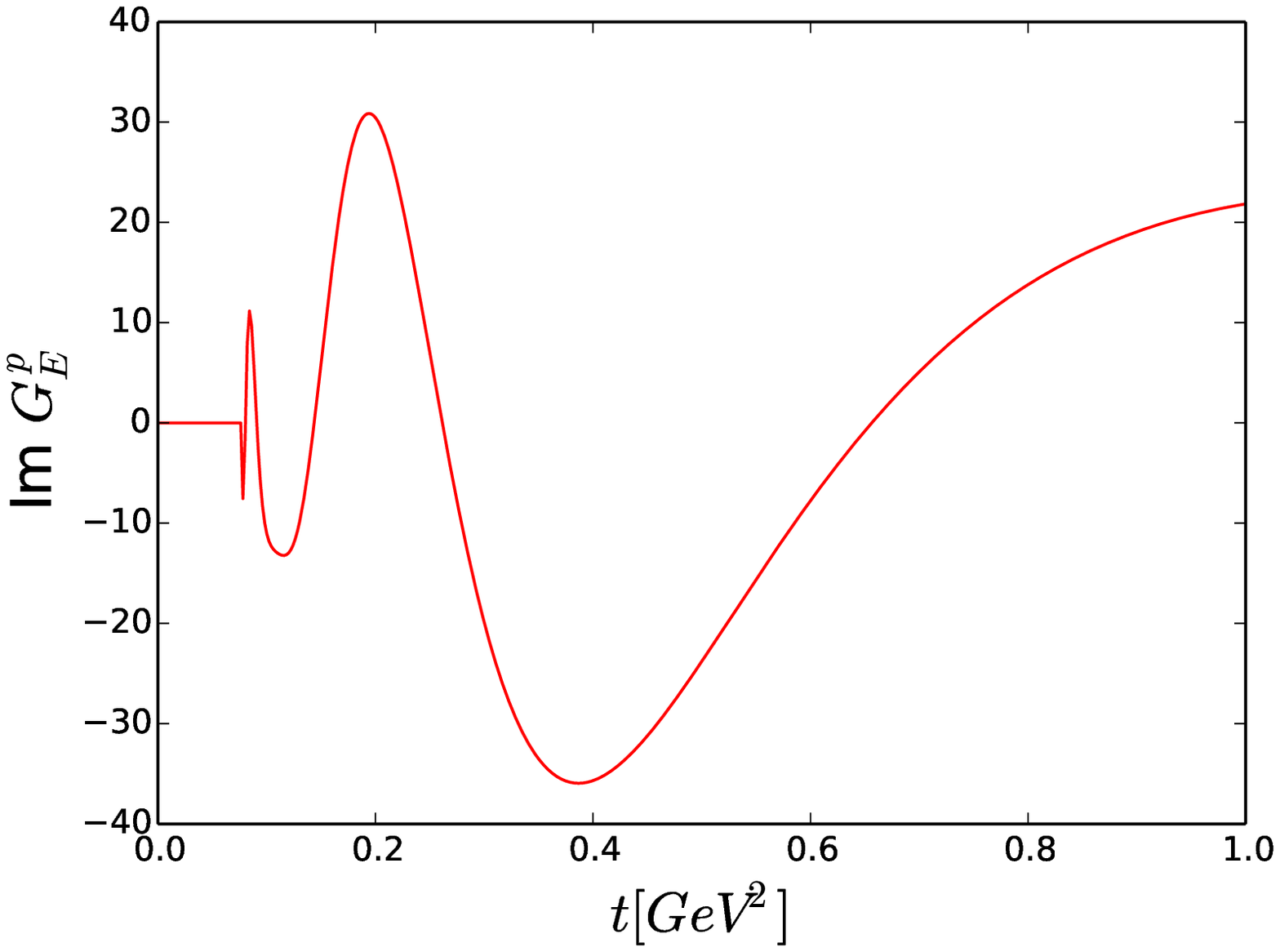}\hglue1mm
\includegraphics[width=0.5\textwidth]{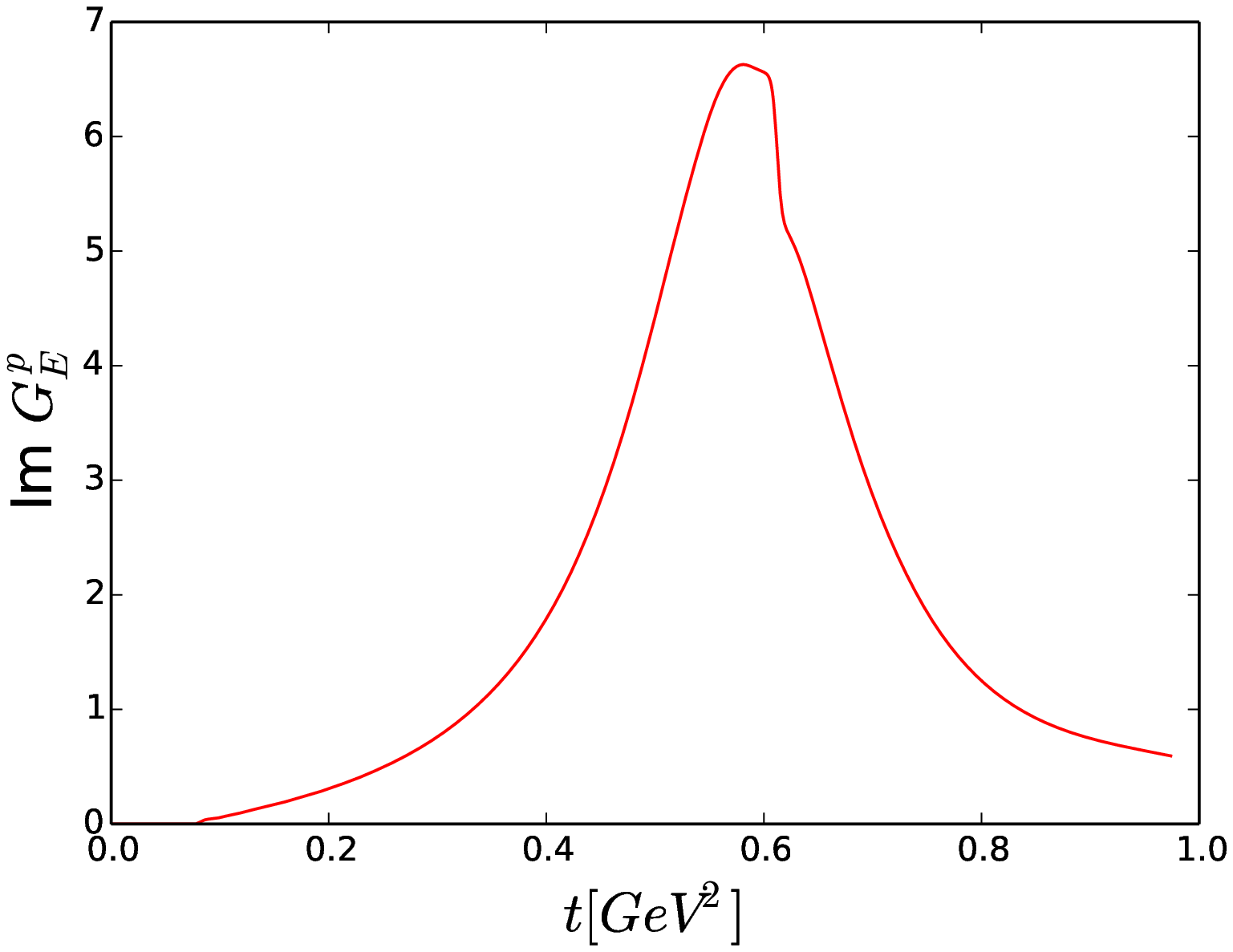}
\caption{Left panel: The spectral function generated by the constrained $z$-expansion fit, as given in Tab.~\ref{table:radii}. Right panel: The spectral function expected from unitarity as fulfilled in Eq.~\eqref{2pi}.}\label{fig:spec}
\end{figure}
Specifically, we consider here four different form factor parametrizations, given in the first column of Tab.~\ref{table:radii}. The second column lists the corresponding fit results to only the MAMI cross sections. Column three shows the results of the fits to the combination of MAMI data and world e-p cross sections. For each result, we quote the electric and magnetic radius and the $\chi^2$ per degrees of freedom (ndf).\newline
The first parametrization is an unconstrained $z$-expansion, Eq.~\eqref{expa}, as we have used in Ref.~\cite{Lorenz:2014vha}. In contrast to that work, here we explicitly include the normalization and point out the related uncertainty for such unphysical fits. We fit nine parameters per form factor and refit all 31 MAMI normalization parameters. The result is a $\chi^2$ value that is to our knowledge lower than in any other published fit. Both electric and magnetic radius from this are far from any previous values. However, a probabilistic interpretation of this fit is hampered by the lack of complete knowledge on the statistical uncertainties. We give the statistical details of this fit in Appendix A. Further details are given there on the error scaling procedure that has been performed by the A1 collaboration on their published data.\newline
The fit in the second row in Tab.~\ref{table:radii} is based on the same parametrization as before but includes bounds on the coefficients. We note an increase of the $\chi^2$/ndf and simultaneously, of the electric radius, but a decrease in the magnetic one. The spectral function from this fit, shown in the left panel of Fig.~\ref{fig:spec}, illustrates why we reject the extrapolation beyond the data for such a fit as unphysical. If we proceed similar to Ref.~\cite{Hill:2010yb}, include the $2\pi$ continuum and raise the $z$ expansion cut to the isoscalar threshold at $9M_{\pi}^2$, we still obtain unrealistic oscillations in the spectral function far below $40M_{\pi}^2$, where we can exclude them, similar to Fig.~\ref{fig:spec}.\newline
In contrast, the full inclusion of the physical constraints from the spectral function as shown in the right panel of Fig.~\ref{fig:spec} is realized in the dispersion relation fit. As given in the third row of Tab.~\ref{table:radii}, this increases the $\chi^2$ further. If we consider an error scaling analogously to the A1 collaboration for a dispersion relation fit instead of a spline fit (see Appendix A), we obtain here a $\chi^2$/ndf of 1.1.\newline
We do not find further improvement by including more effective pole terms above $t=40M_{\pi}^2$. At this point, the deterioration of the fit for fully included constraints could still be related to unknown continua above $t=40M_{\pi}^2$ that are not well described by pole terms. Therefore, we perform a new conformal mapping with $t_{cut}=40M_{\pi}^2$ and add an expansion in the new variable to the above dispersion relation fit. Numerically, we only find a convergence of the new fit, if we fix the normalization parameters of the MAMI data which would correspond to a slightly smaller $\chi^2$/ndf. In absolute values, it remains unaffected on the percent level showing no improvement for this combined approach of dispersion relations and conformal mapping. The results of this attempt are given in row 4 of Tab.~\ref{table:radii}.\newline
For the given results, the $\chi^2$ is calculated with a diagonal covariance matrix as suggested by the A1 collaboration. Alternatively, we checked whether the inclusion of the systematic errors in the covariance matrix improves the fits. Partly, we also replaced the normalization parameters by the corresponding uncertainties in the covariance matrix, see Appendix B for details. However, this procedure does not improve the fits significantly.\newline
All cross sections are corrected by our TPE calculations. As we have seen in Fig.~\ref{fig:app}, at the kinematics of the MAMI data the contribution $\delta_{2\gamma,\Delta}$ is significantly smaller than $\delta_{2\gamma,N}$. For the higher $Q^2$ values occurring in the other cross sections, $\delta_{2\gamma,\Delta}$ is more relevant. However, regarding the fits, the influence of the parametrization is much stronger than the influence of the included radiative corrections. We evaluate this influence mainly from the $\chi^2$ values of the fits and the corresponding extracted electromagnetic radii.\newline
The world data compilation on e-p scattering that is partly included in the fits, consists of 23 individual data sets itself. The treatment of their normalization uncertainty adds a source of ambiguity to these fits. Some authors fit the normalization for each set freely, others insist on a fixed data normalization, despite the large uncertainty of at least several $\%$ \cite{Sick14}. However, compared to the choice of the data set, the normalization treatment in the world data set is negligible. The values given here consider freely fitted normalizations in the world data set.\newline
For definiteness, we explicitly show the results of the pure DR fit to the MAMI data in Fig.~\ref{fig:drc}. The parameters of this fit are listed in Tab.~\ref{table:drcvalues}. Important for the comparison to the data is the inclusion of the normalization parameters, here applied to the cross sections to shift them accordingly. In order to illustrate the composition of the data from the different spectrometers, we include a close-up of the data sets in Appendix C. This also shows that the deviations of the fit from the individual data sets are approximately Gaussian distributed, which would allow an error scaling as described in Appendix A. The prediction from this DR fit for the form factor ratio is given in Fig.~\ref{fig:droff}. In comparison we show recent measurements from Jefferson Laboratory using recoil polarization techniques that yield the form factor ratio directly \cite{Ron:2011rd, Zhan:2011ji}, but were not fitted here. Below $Q^2 \simeq 0.2$~GeV$^2$, unphysical fits tend to produce oscillations in the magnetic form factor, clearly visible in the ratio \cite{Bernauer10, Lorenz:2014vha}. Full constraints lead to the disappearance of the oscillations, as shown in Fig.~\ref{fig:droff}.  
\begin{table}[ht!]
\centering
\begin{tabular}{|l|c|c|c||l|c|c|c|}
\hline
$V_{isoscalar}$ & $m_V$ & $a_1^V$ &$a_2^V$ & $V_{isovector}$ & $m_V$ & $a_1^V$ & $a_2^V$  \\
\hline
$\omega$ & $0.783$	& $0.500$	& $-0.190$	& $v_1$ & $2.330$	 &$-1.911$ & $0.314$	\\
$\phi$   & $1.019$	& $0.375$	& $-0.861$	& $v_2$ & $2.192$	 & $0.644$ & $2.265$	\\
$s_1$    & $3.052$	&$-0.446$	& $ 0.512$	& $v_3$ & $4.272$	 & $0.173$ & $-0.322$	\\
$s_2$    & $1.571$	& $0.095$	& $-2.388$	& $v_4$ & $2.454$	 & $0.158$ & $0.064$	\\
$s_3$    & $2.580$	& $0.760$	& $-0.538$	& $v_5$ & $2.492$	 & $0.142$ & $-0.372$	\\
\hline
\end{tabular}
\begin{tabular}{|l|l|l|l|l|l|l|l|l|l|l|l||l|l|l|l|}
 \hline
 n1 & $0.9982$ & n5 & $1.0059$ & n9  & $1.0056$ & n13 & $1.0052$ & n17 & $1.0008$ & n21 & $0.9995$ & n25 & $1.0056$ & n29 & $1.0078$ \\
 n2 & $0.9928$ & n6 & $1.0017$ & n10 & $1.0025$ & n14 & $1.0044$ & n18 & $1.0076$ & n22 & $0.9975$ & n26 & $1.0080$ & n30 & $0.9985$ \\
 n3 & $1.0051$ & n7 & $1.0003$ & n11 & $1.0000$ & n15 & $1.0023$ & n19 & $1.0055$ & n23 & $0.9996$ & n27 & $1.0004$ & n31 & $1.0060$ \\
 n4 & $1.0078$ & n8 & $0.9975$ & n12 & $1.0031$ & n16 & $1.0006$ & n20 & $1.0035$ & n24 & $0.9984$ & n28 & $1.0011$ &     &  \\
 \hline
\end{tabular}
\caption{The parameters obtained from the DR approach fit to the MAMI data: 
regular (upper panel) and normalization parameters (lower panel). 
The latter have to be multiplied to the cross sections to allow for a meaningful comparison to the data, cf. Eq.~\eqref{chi}. The normalizations are assigned to the data as given in \cite{Bernauer10}. Masses $m_V$ are given in GeV and couplings $a_i^V$ in GeV$^{2}$.}
\label{table:drcvalues}
\end{table}
Due to the lack of a probabilistic interpretation of the $\chi^2$ values in our case, common methods like an ellipse in the $\chi^2$ landscape around the minimum fail here. Instead, we choose a bootstrapping procedure for a further error estimation of individual fits. For the DR approach fit to the MAMI data, we obtain the following $3\sigma$ uncertainties:
\begin{align}
r_E^p = 0.840~(0.828-0.855) \text{fm}~,\\ 
r_M^p = 0.848~(0.843-0.854) \text{fm}~.
\end{align}
For the exact error procedure, see Appendix D. Despite the improvements performed here, these values agree well within their errors with previous dispersion relation fits \cite{Lorenz:2012tm, Belushkin:2006qa}. Regarding the radius extraction, the TPE corrections are much less relevant than the inclusion of constraints on the FFs. Specifically, for the parametrizations in Tab.~\ref{table:radii}, we find fits to the uncorrected cross sections with changes in the radii of less than 1$\%$. This also holds for an inclusion of only the nucleon-TPE. Due to this insensitivity of the radii to the corrections, we refrain from another iteration of the TPE calculation with NFFs extracted from the corrected data.

\section{Muonic hydrogen and the third Zemach moment}
\noindent As mentioned earlier, the proton charge radius is also relevant in QED calculations of atomic energy splittings, like e.g. the Lamb shift. This partly depends not only on the proton radius but also on higher moments of the charge distribution, the Zemach moments $\langle r^n\rangle_{(2)}$. Following an overview over such calculations for hydrogen like atoms \cite{Eides:2000xc}, or more specifically Ref.~\cite{Friar:2005jz}, one can consider the proton structure contribution to the hydrogen Lamb shift at leading $\mathcal{O}(\alpha^4)$ and $\mathcal{O}(\alpha^5)$ as
\begin{align}
 \Delta E = \frac{2\pi\alpha}{3}|\phi_n(0)|^2\left(\langle r_E^2\rangle - \frac{m_r\alpha}{2}\langle r^3\rangle_{(2)} + \ldots\right),
\end{align}
since $|\phi_n(0)|^2$, the wave function of the $n$th $S$-state at the origin, contains $\alpha^3$. $m_r$ is the reduced mass of the lepton-proton system, showing a larger relative impact of the second term for muonic than for regular hydrogen. Disagreements between several field theoretical calculations of the $\mathcal{O}(\alpha^5)$ corrections to the Lamb shift in muonic hydrogen, see below, motivate our phenomenological determination of the third Zemach moment in this work.\newline
Zemach originally calculated the hyperfine shift in hydrogen and for this aim introduced a convolution of the electric and magnetic distributions \cite{Zemach:1956zz}. In later determinations of the Lamb shift, higher moments of the charge distribution also went under the name ``Zemach moment'' \cite{Friar:1978wv, Friar:2005jz}:
\begin{equation}
 \langle r^n\rangle_{(2)} = \int d^3r r^n\rho_{(2)}(r)
\end{equation}
where 
\begin{align}
 \rho_{(2)}(r) &= \int d^3r'\rho_E(\bm{r'})\rho_E(|\bm{r'-r}|) = \int \frac{d^3q}{(2\pi)^3}e^{-i\bm{qr}}G_E^2(\bm{q^2}).
\end{align}
In the limit of a proton mass that dominates the system, it was shown \cite{Pachucki:1996zza} that one can approximate the third Zemach moment as
\begin{align}
 \langle r^3\rangle_{(2)} = \frac{48}{\pi}\int_0^{\infty}\frac{dq}{q^4}\left(G_E^2(q^2)-1+\frac{q^2\langle r^2_E\rangle}{3}\right).\label{zemach}
\end{align}
The terms that are neglected in this limit correspond to higher-order recoil corrections \cite{Eides:2000xc,Pachucki99}. The larger sensitivity was initially used as an argument for a Lamb shift measurement in muonic hydrogen \cite{Pachucki:1996zza}. After these measurements, it was debated whether the third Zemach moment could solve the radius discrepancy \cite{DeRujula:2010dp, DeRujula:2010ub, DeRujula:2010zk}. For some FF parametrizations without physical motivation that were fitted to older data, the latter solution was shown to be improbable \cite{Cloet:2010qa}. However, the impact of more realistic FFs that agree with physical constraints and fully radiatively corrected data on Eq.~\eqref{zemach} was called for. Field theoretically, the $\mathcal{O}(\alpha^5)$-contribution has been considered in \cite{Pachucki99, Pachucki:1996zza, Faustov:1999ga, Carlson:2011zd, Birse:2012eb, Hill:2011wy, Alarcon:2013cba}, using and pointing out different approximations and yielding partly deviating results. A recent calculation in heavy baryon chiral perturbation theory including the $\Delta$-resonance \cite{Peset:2014jxa, Peset:2014yha} found only for the sum of inelastic and elastic contribution a similar value to some of the previous works, but large deviations for the individual parts.\newline
The DR-fits yield for the third Zemach-moment the following $3\sigma$-bootstrap-errors:
\begin{align}
 \langle r^3\rangle_{(2)} = (1.3 - 3.8)\text{fm}^3
\end{align}
Using this upper limit in the Lamb-shift calculation for muonic hydrogen would shift the proton charge radius
\begin{align}
 r_E^p = (0.841 \rightarrow 0.843)\text{fm}.
\end{align}
Thus, the discrepancy between regular electronic and muonic hydrogen is largely untouched by our results.

\section{Discussion and Conclusions}
\noindent In the first part of this work, we have explicitly calculated the two-photon corrections to electron-proton scattering including nucleon and $\Delta$ intermediate states using phenomenological information on the vertices. In particular, we have analysed the main uncertainties in these calculations. On the one hand, we varied the $\gamma N\Delta$-vertices from a previous implementation by including the Coulomb contribution and updating the photocoupling values. Alternatively, we also employed data on the $Q^2$ dependence of the nucleon-$\Delta$ transition from electroproduction of nucleon resonances in terms of helicity amplitudes. On the other hand, we found that the dominating uncertainty is based on the choice of the NFFs in the $1\gamma$ amplitude that enters the cross section correction. This dependence is what we expect analytically from the partial cancellation of the FF dependence in the TPE correction Eq.~\eqref{tpe}. Numerically, we show that this dependence leads to deviations of more than a factor 10 to a previous calculation for some kinematics given there. We apply our TPE calculation to the MAMI cross sections on e-p scattering \cite{Bernauer10}, where the kinematical conditions lead to a much smaller $\Delta$ contribution $\delta_{2\gamma,\Delta}$ compared to the pure nucleon contribution $\delta_{2\gamma,N}$. In contrast to this, for example at $Q^2=3~$GeV$^2$, the contributions from an excited intermediate state are of the same order as the elastic contribution.\newline 
The second part of this work deals with the determination of the elastic electromagnetic NFFs. For this aim, we perform fits to the corresponding scattering cross sections. Regarding the data treatment, our work has several advantages towards most other analyses.  Specifically, we employ our full TPE corrections, instead of old approximations with partly the wrong sign. Moreover, we use the cross sections directly instead of the FFs extracted from the latter via a Rosenbluth separation, which would induce further systematic errors. However, the most relevant advantage of our fits is the inclusion of the full physically motivated constraints from analyticity and unitarity. We show how their subsequent consideration influences the fits. Starting from a flexible fit function based on a conformal mapping without constraints, we can describe the data perfectly in a statistical sense. But we show explicitly, that even rough constraints on such a function still lead to an unrealistic spectral function. Thus we turn to a dispersion relation approach to include the full mass-related information from the spectral function. We want to emphasize that the extrapolation from the lowest data points to the origin corresponds to an uncertainty that we expect to be biased. Any curvature in the real FF below the given data obviously leads to a bias due to the missing data. This might explain why $r_E^p$ in conventional fits tends to come out larger, also in statistically sophisticated analyses \cite{Graczyk:2014lba}.\newline
In the third part, we have determined the third Zemach moment from our form factor fits. Note that for this calculation, the form factors are relevant beyond the data range of the corresponding MAMI measurement. Thus it was crucial to include further cross sections, also in a region where the TPE corrections including the $\Delta$ become more relevant. However, constraining the asymptotic behavior of the NFFs according to quark counting rules as in Ref.~\cite{Lorenz:2012tm} has a similar effect on the extracted Zemach moment. The remaining discrepancy between the radius values from ordinary and muonic hydrogen remains largely unaffected by our results. Measurements of hydrogen energy splittings, see \cite{Beyer:2013daa}, lead to proton charge radii that are 1-2$\sigma$ off the small value obtained in our physical fits. However, the uncertainties of the spectroscopic radius determination is under debate \cite{Karshenboim:2014baa}.
\bigskip\newline
\textbf{Acknowledgements}\newline
\noindent The authors are grateful for useful discussions with S.~Karshenboim, J.~Bernauer, T.~Udem, R.~Gilman, G.~Paz, G.~Lee and C.~Carlson. I.~Sick is additionally acknowledged for providing his world data compilation. The authors thank D.-Y.~Chen for initial hints on numerical issues and W.~Wang for checking the TPE code.
Further, A.~De~Rujula's suggestion to have a look at the third Zemach moment from realistic FFs is acknowledged.
One of the authors (I.T.L.) wants to thank the Institute of High Energy Physics, Beijing, for its hospitality and support. 
This work is supported in part by the DFG and the NSFC through
funds provided to the Sino-German CRC 110 ``Symmetries and
the Emergence of Structure in QCD'', the NSFC Grants 
No.~10975146, No.~11035006 and No.~11475192 and by the Helmholtz Association under contract HA216/EMMI.

\section{Appendices}

\subsection{Statistics and constraints}
\begin{figure}[t]
\centering
\includegraphics[width=0.78\textwidth]{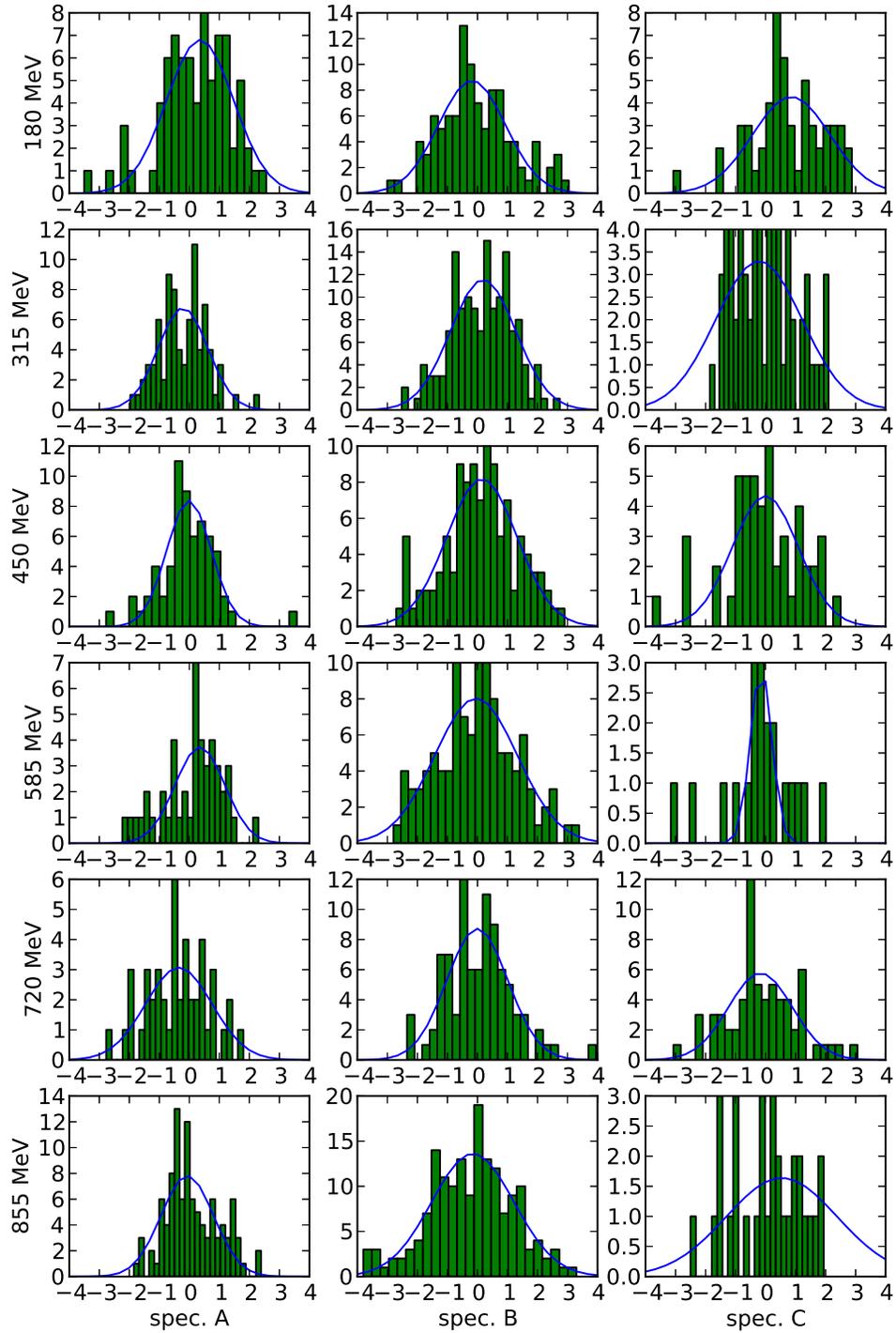}
\caption{The distributions of the deviations between a DR-fit and the individual MAMI data sets, that correspond to six different energy settings of the incoming electron beam and three different spectrometers.\label{fig:widths}}
\end{figure}

\noindent The improved radiative corrections also interfere with the normalization. As recommended by the A1 collaboration \cite{ Bernauer10}, after the inclusion of our calculated corrections we refit the normalization parameters using a flexible fit function. This is suggested in order to adjust the different data sets relative to each other by maximizing their overlap. Here, we compare their suggestion to use a polynomial for this procedure to the use of an unconstrained $z$ expansion as used in \cite{Lorenz:2014vha}. Both alternatives give variations of the normalization parameters in the range of 1$\%$. Here, a few words on statistics are at hand. This normalization procedure gives for the polynomial an absolute $\chi^2$ of 1563, as for the uncorrected data. For a $z$ expansion with the normalization parameters from a spline fit, the $\chi^2$/ndf is the same again. However, with fitted normalizations, the absolute $\chi^2$ is further reduced to 1537. Considering this result from a purely statistical standpoint put forward for example in \cite{Bernauer10}, one should prefer this fit towards any of those given in the same reference due to the small $\chi^2$ value. In terms 
of the one-sided p-value
\begin{equation}
 p(\chi^2, ndf) = \frac{\Gamma(ndf/2, \chi^2/2)}{\Gamma(ndf/2)}
\end{equation}
this fit gives an acceptable data description with at least $p(1537,(1422-51)) \approx 0.1$. The ``best'' value published so far for fits to these data \cite{Bernauer10} amounts to $p(1563,(1422-51)) \approx 0.04$. However, due to frequent misunderstandings of this basic fact, we point out again that a strict probabilistic interpretation of $\chi^2$ values is only valid if the errors are well known. This is not the case for the Mainz A1 data where several effects that contribute to the statistical error cannot be quantified a priori. According to Ref.~\cite{Bernauer10}, chapter 8, these effects include the normalization to the luminosity measurement, the uncertainty of the current measurement for the 315~MeV data, the statistical error of the background estimation and undetected slight variations of the detector and accelerator performance. In the A1 analysis \cite{Bernauer10}, their size is approximated under the assumption that the deviations of the data from a certain spline fit follow a Gaussian distribution. Specifically, these deviations are considered individually for each of the 18 data sets (six energies, three spectrometers). The widths of the distributions of the deviations are used to scale the errors. This is carried out iteratively until a $\chi^2$ close to 1 is obtained. While this might be the best available approximation of the errors in this case, it results in their statistical meaning vanishing.\newline
According to the Weierstrass approximation theorem, in general, a polynomial with a sufficient number of degrees of freedom can fit any curve. Therefore the polynomial fit used in the Mainz-A1 error scaling procedure might describe the data too precisely and thus be nonsuitable to estimate the unknown errors. The distributions of the deviations between a DR-fit and the individual MAMI data sets, as shown in Fig.~\ref{fig:widths}, would allow us to perform an error-scaling for this fit analogous to the A1 procedure. However, unless specified otherwise, we refrain from an additional error scaling in this work. 

\subsection{Fitting procedure}
\noindent In contrast to the original MAMI-analysis, we partly consider the inclusion of the correlated errors in the fitting procedure, by minimizing the $\chi^2$ function 
\begin{align}
 \chi^2 = \sum_k(n_k C_i - C(Q^2_i,\theta_i,\vec{p}\,))[V^{-1}]_{ij}(n_k C_j - C(Q^2_i,\theta_i,\vec{p}\,))\label{chi}
\end{align}
where $C_i$ are the cross section data at the points $Q^2_i,\theta_i$ and $C(Q^2_i,\theta_i,\vec{p}\,)$ are the cross sections for a given FF parametrization for the parameter values contained in $\vec{p}$. The covariance matrix is given by
\begin{align}
 V_{ij} = \sigma_i\sigma_j\delta_{ij} + \nu_i\nu_j~,
\end{align}
where $\sigma_i$ are the statistical and $\nu_i$ the systematical errors, here uncorrelated and correlated, respectively. Note that the correlated errors considered here have not been given in the original publication but only in the later online version. For the inclusion of the normalization uncertainty, we perform two alternative methods, once via free fit parameters $n_k$ in Eq.~\eqref{chi} and once via their treatment as completely correlated systematical errors in the covariance matrix. In principal, these methods were shown to be equivalent \cite{Demortier99}.

\newpage
\subsection{Cross sections and Dispersion Relation Fit}
\begin{figure}[h!]
\centering
\includegraphics[width=0.5\textwidth]{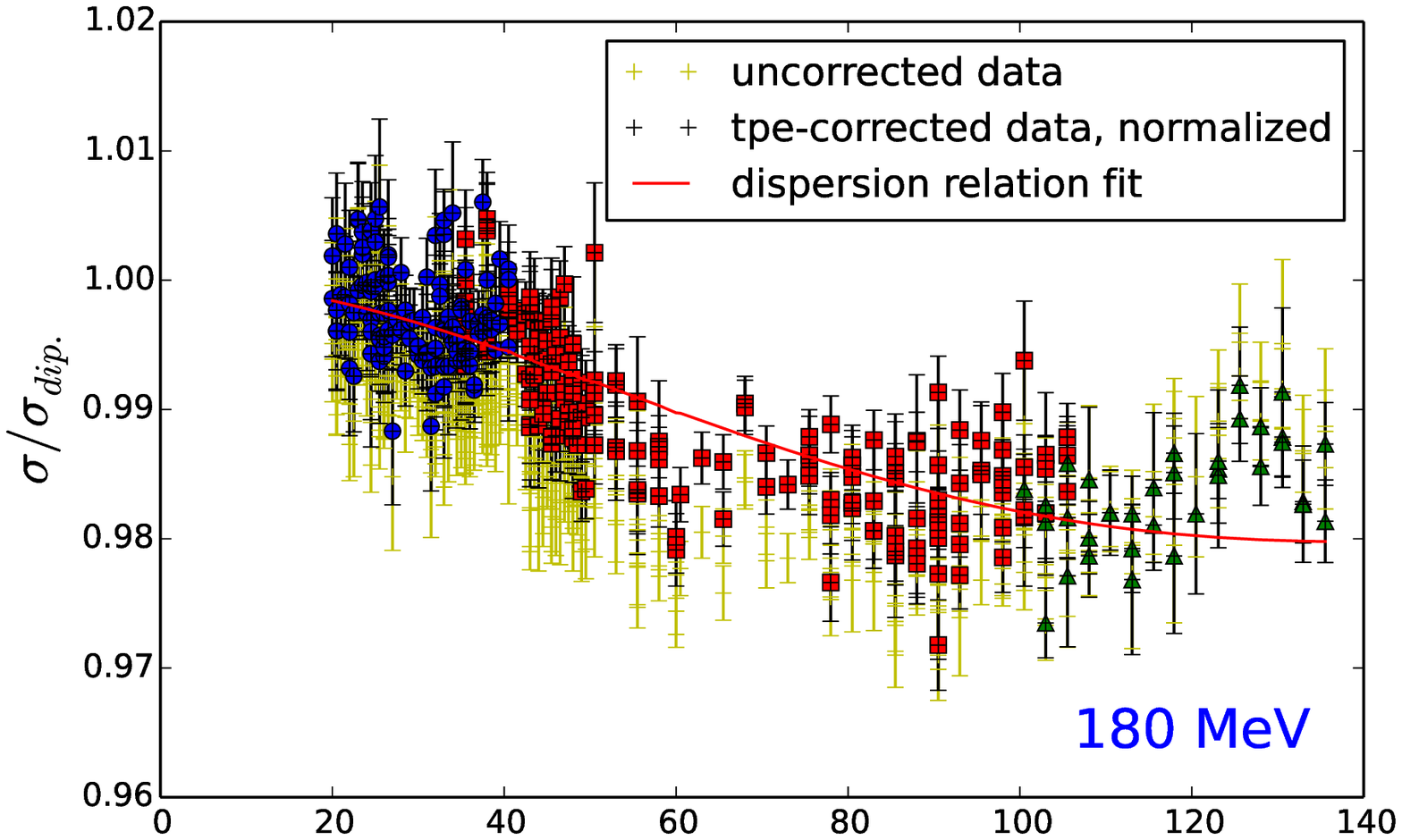}\hglue1mm
\includegraphics[width=0.5\textwidth]{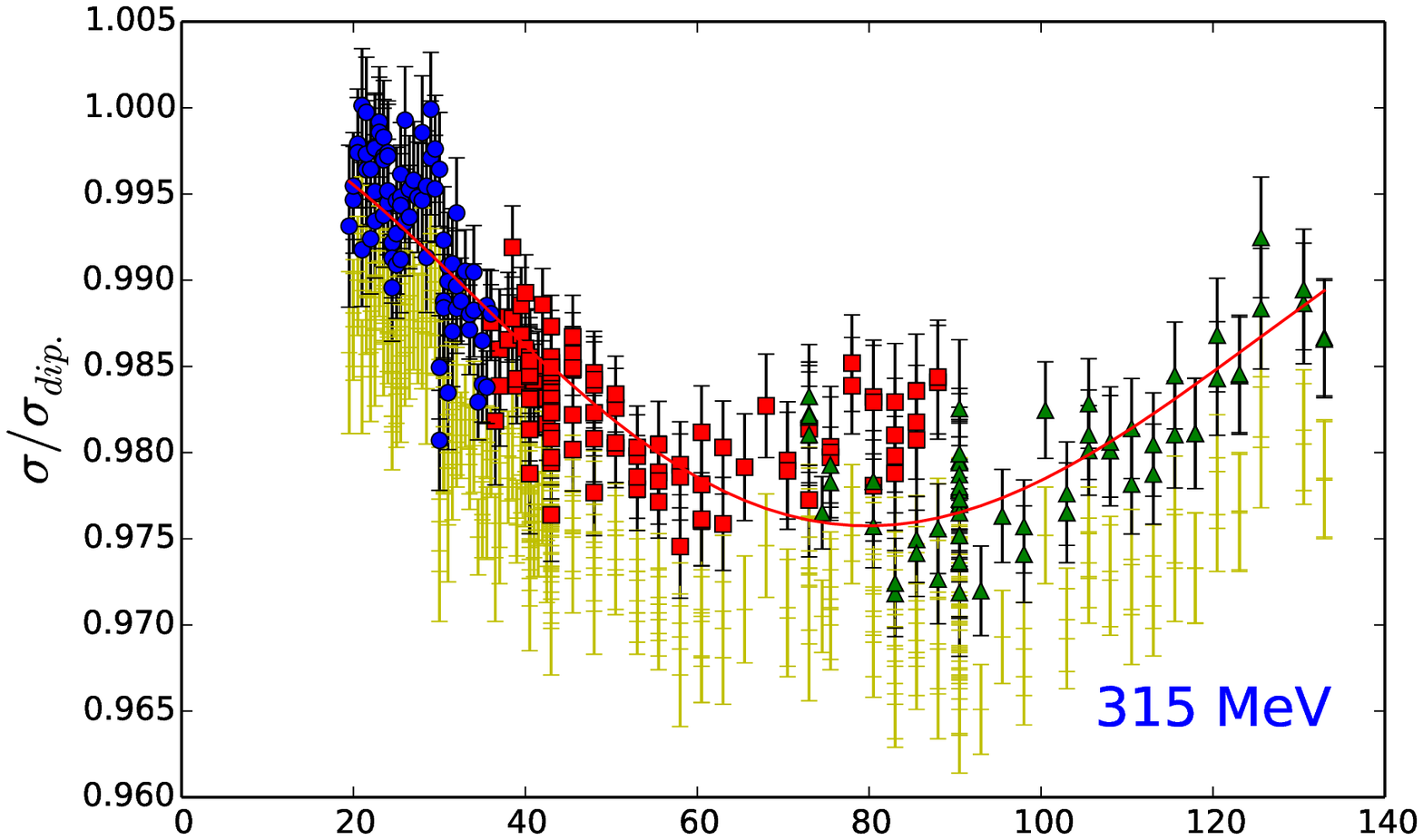}\vglue1mm
\includegraphics[width=0.5\textwidth]{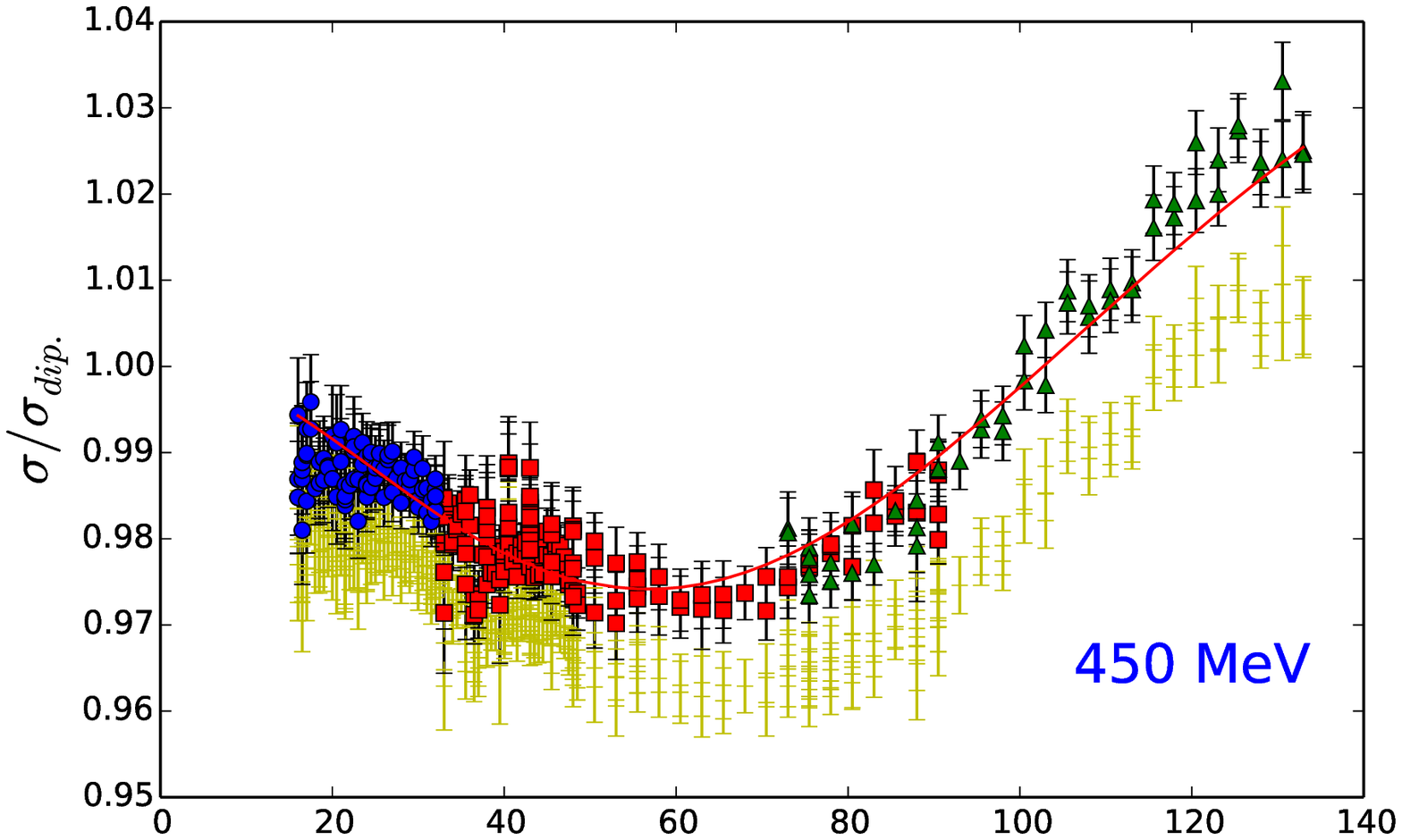}\hglue1mm
\includegraphics[width=0.5\textwidth]{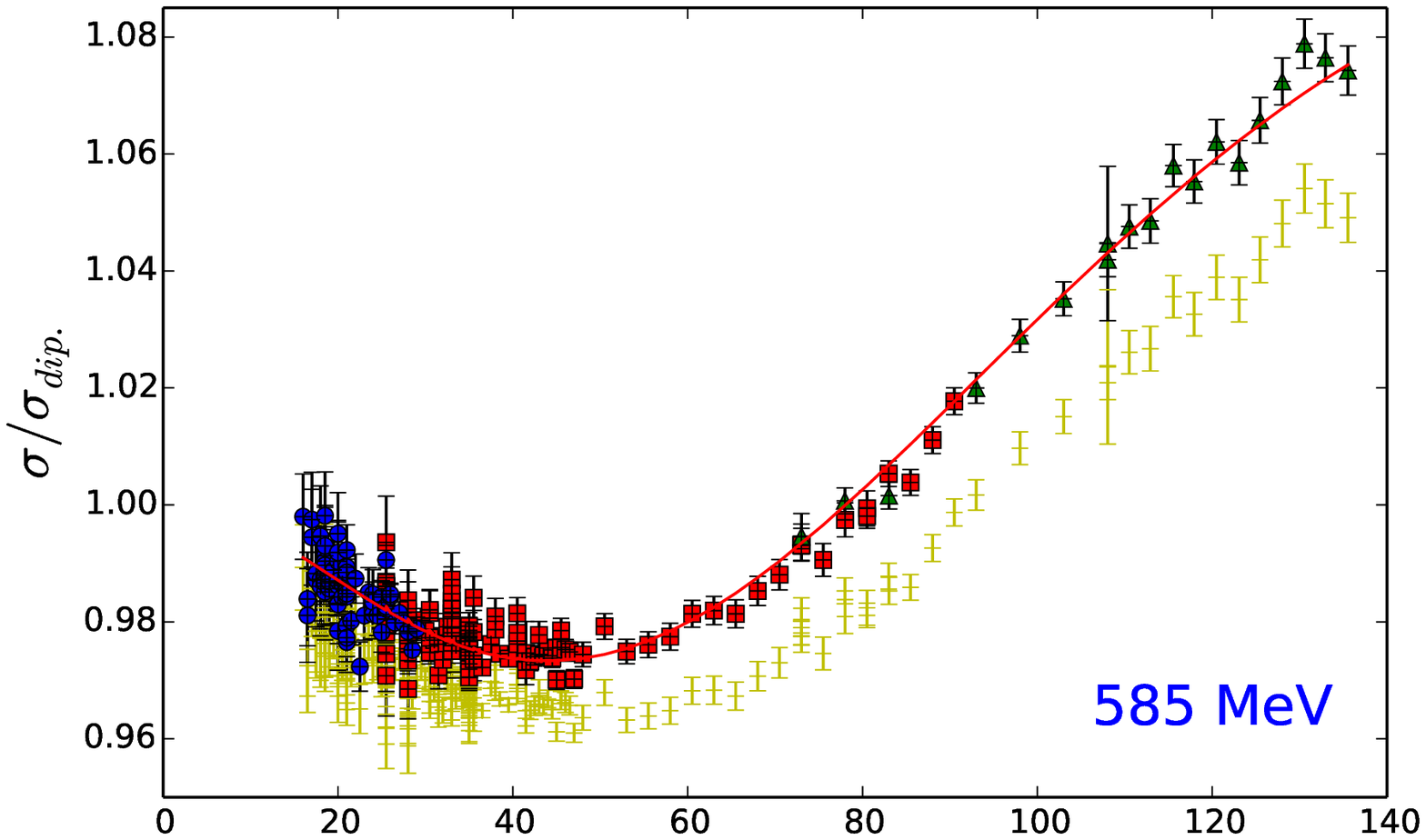}\vglue1mm
\includegraphics[width=0.5\textwidth]{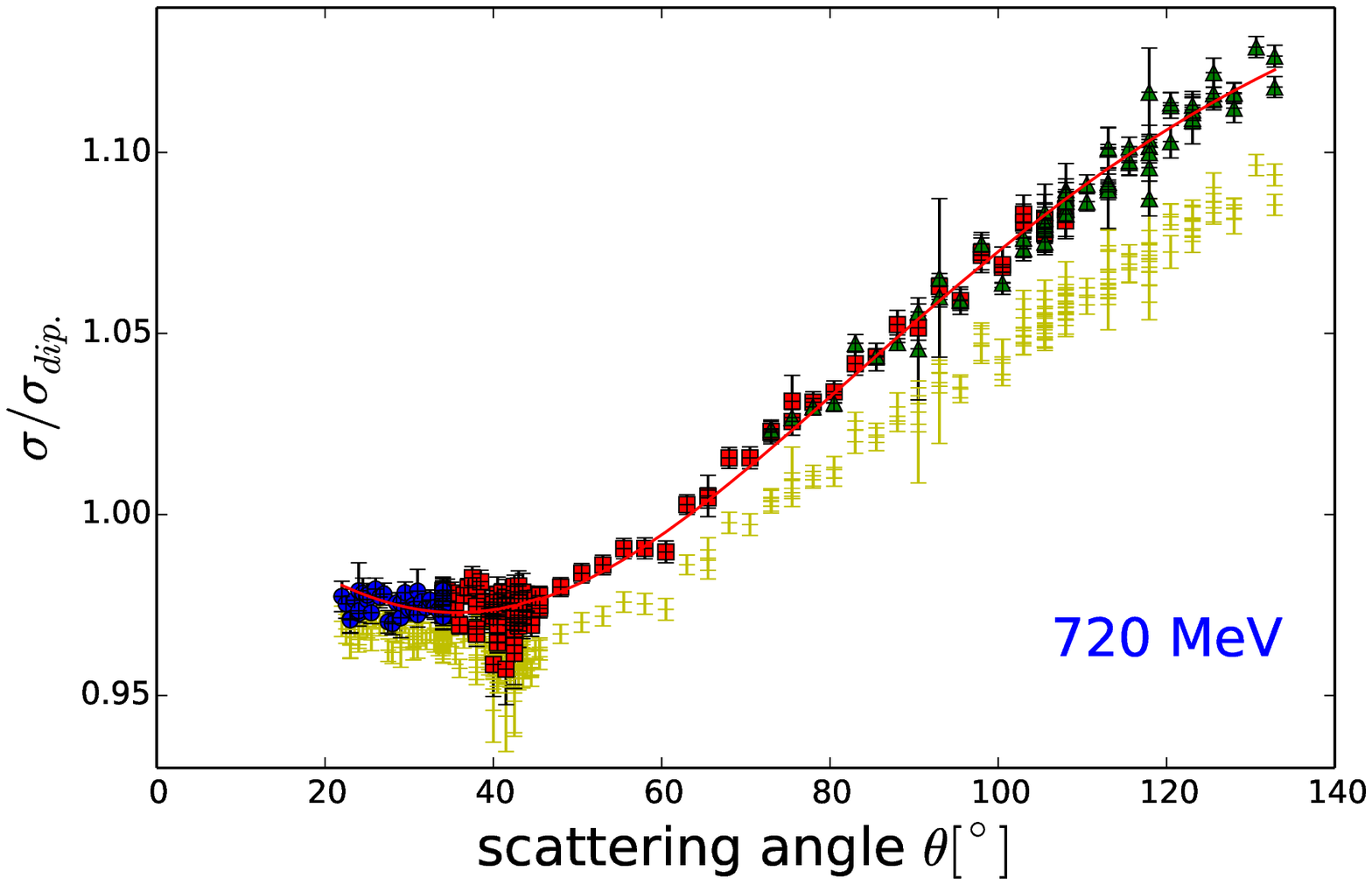}\hglue1mm
\includegraphics[width=0.5\textwidth]{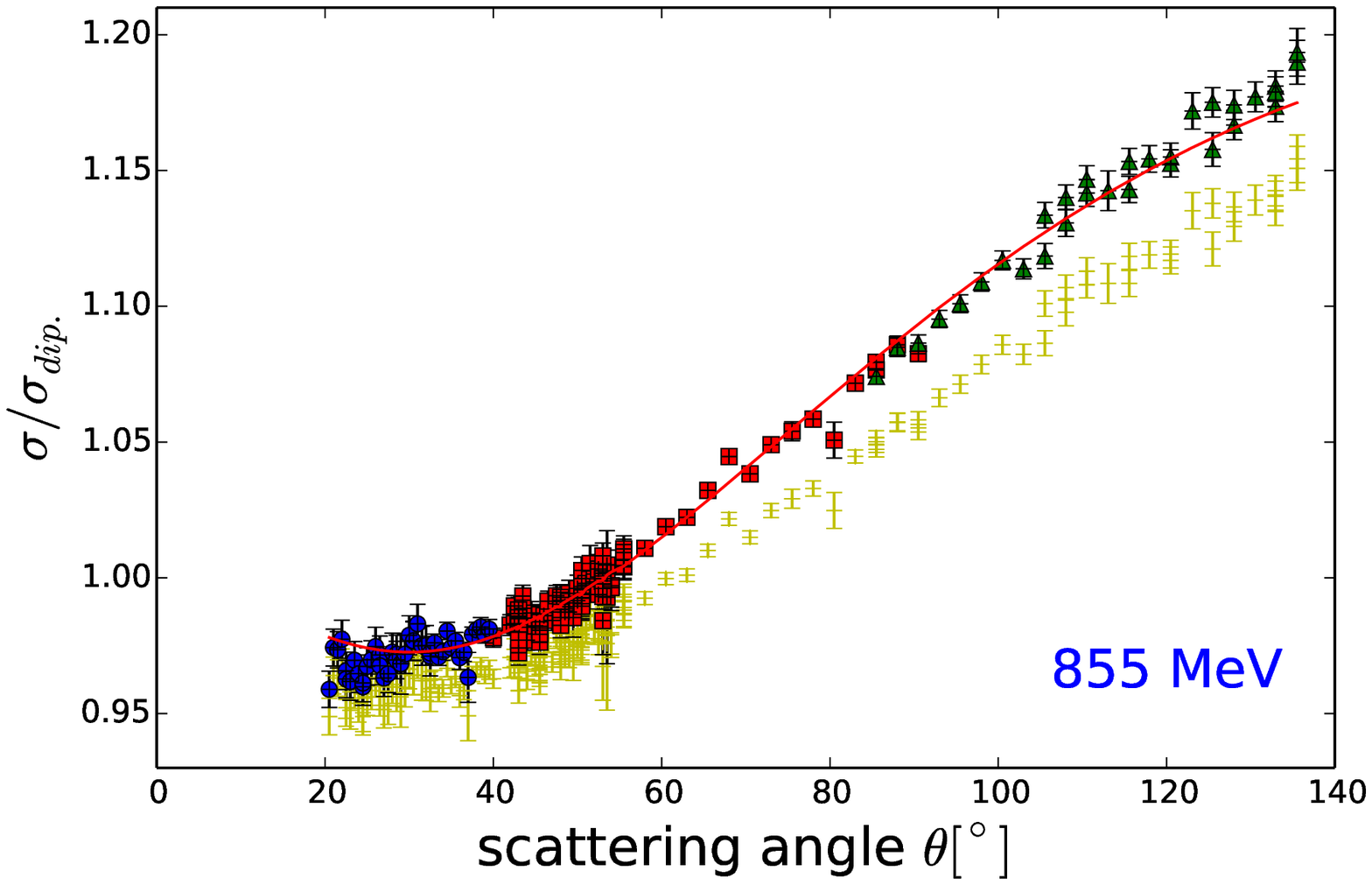}
\caption{Results from the DR approach as given in Tab.~\ref{table:drcvalues}. Close-up of the cross sections with specified spectrometers A (red, squares), B (blue, circles) and C (green, triangles). Energies of the incoming electron beam given.\label{fig:drfit}}
\end{figure}

See Fig.~\ref{fig:drfit} for a comparison of the cross sections obtained from the dispersion relation fit with two-photon exchange corrected and uncorrected data.
\subsection{Bootstrap procedure}
\noindent To estimate the fit errors for the radii, we use a bootstrap procedure following Ref.~\cite{boot}. We simulate a high number of data sets compared to the number of data points by randomly varying the points in the original set within the given errors assuming their normal distribution. We fit to each of them separately, derive the radius from each fit and consider the distribution of these radius values, which is sometimes denoted as bootstrap distribution. The artificial data sets represent many real samples. Therefore, our radius distribution represents the probability distribution that one would get from fits to data from a high number of measurements. Although standard books on numerics refer to this as the ``quick and dirty'' method, the wide acceptance nowadays can be put on firm statistical ground \cite{numrecipe}. The precondition for using this method are independent and identically distributed data points which is fulfilled in our case, since the $\chi^2$ sum does not depend on the sequential order of the contributing points. For $n$ simulated data sets, the errors thus scale with $1/\sqrt{n}$.

\bibliographystyle{unsrturl}
\bibliography{tpebib}

\end{document}